%% file: byteback.tex
\documentclass[a4papers,runningheads]{llncs}

\usepackage{hyperref}

\newif\ifblind
\blindfalse

\newif\ifdraft
\drafttrue

\newif\iflong
\longfalse

\usepackage[T1]{fontenc}
\usepackage{times}
\usepackage[scaled=0.81]{beramono}

\usepackage{graphicx}

\usepackage{tikz}
\usetikzlibrary{shapes,arrows,positioning,calc,fit,intersections,through}

\definecolor{bbcol}{RGB}{27,158,119}
\definecolor{boogiecol}{RGB}{217,95,2}
\definecolor{grimpcol}{RGB}{117,112,179}
\definecolor{bytecodecol}{RGB}{230,171,2}

\usepackage[inline]{enumitem}
\usepackage{ragged2e}

\usepackage{booktabs}
\usepackage{relsize}
\usepackage{changepage}            
\usepackage[skip=5pt,font=small,strut=off]{caption}
\usepackage[skip=1pt]{subcaption}  

\graphicspath{{./figs/}}
\DeclareGraphicsExtensions{.pdf,.png}

\usepackage{amsmath,amsfonts}
\usepackage{mathtools}
\usepackage{array}
\usepackage{xspace}

\usepackage{pgfkeys,pgfplots,numprint,relsize,refcount}
\xspaceaddexceptions{\%}  
\def\NAN{??}              
\def\keyfamily{/bb}

\usepackage[firstpage]{draftwatermark}

\input{experiments.tex}

\input{stats.tex}

\DeclareDocumentCommand{\n}{t. t: o m o O{} t|}{%
  \begingroup
  \pgfkeys{/pgf/fpu=true}
  \IfBooleanTF{#2}{%
    \pgfkeyssetvalue{/tmp/value}{#4}%
    \pgfkeyssetvalue{/tmp/found}{found}%
  }{
    \pgfkeysifdefined{\keyfamily#4}{%
      \pgfkeyssetvalue{/tmp/value}{\pgfkeysvalueof{\keyfamily#4}}%
      \pgfkeyssetvalue{/tmp/found}{found}%
    }{}
  }%
  \pgfkeysifdefined{/tmp/found}{%
    \IfNoValueF{#5}{%
      \pgfkeyssetvalue{/tmp/multiplier}{#5}%
      \pgfmathparse{\pgfkeysvalueof{/tmp/multiplier} * \pgfkeysvalueof{/tmp/value}}%
      \pgfkeyslet{/tmp/value}\pgfmathresult%
    }%
    \IfBooleanTF{#1}{%
      \pgfkeysvalueof{/tmp/value}%
    }{
      \IfNoValueTF{#3}{%
        \pgfmathprintnumber%
        [set thousands separator={\,},int detect,#6]%
        {\pgfkeysvalueof{/tmp/value}}%
      }{
        {\pgfmathprintnumber%
          [precision=#3,fixed,zerofill,set thousands separator={\,},#6]%
          {\pgfkeysvalueof{/tmp/value}}}%
      }%
    }%
    \IfBooleanT{#7}{{\smaller[1.2]\%}}%
  }{
    \NAN%
  }%
  \pgfkeys{/pgf/fpu=false}
  \endgroup%
}

\DeclareDocumentCommand{\exref}{m o}{%
  \begingroup%
  \IfNoValueTF{#2}{%
    \pgfkeyssetvalue{/tmp/result}{\getrefnumber{#1}}%
  }{%
    \pgfkeyssetvalue{/tmp/from}{\getrefnumber{#1}}%
    \pgfkeyssetvalue{/tmp/to}{\getrefnumber{#2}}%
    \pgfmathparse{1 + \pgfkeysvalueof{/tmp/to} - \pgfkeysvalueof{/tmp/from}}%
    \pgfkeyslet{/tmp/result}\pgfmathresult%
  }%
  \pgfmathprintnumber[int detect,assume math mode=true]{\pgfkeysvalueof{/tmp/result}}%
  \endgroup%
}

\usepackage{listings}

\lstdefinestyle{displayed}{
  numbers=left, %
  firstnumber=last, 
  frame=single, %
  breaklines=true,
  numbersep=2pt,
  frame=tb,
  numberstyle=\tiny,
  tabsize=2,
  captionpos=b,
  xleftmargin=0mm, %
  xrightmargin=0mm, %
  basicstyle=\ttfamily\scriptsize,
  keywordstyle=\bfseries\ttfamily,
  commentstyle=\color{darkgray}\itshape\ttfamily,
  keepspaces=true,
  columns=fixed,
  escapeinside={(*}{*)},
  mathescape=true,
  showstringspaces=false
}

\lstdefinestyle{plain}{
  numbers=none, %
  frame=none, %
  breaklines=false,
  tabsize=2,
  xleftmargin=2mm, %
  xrightmargin=2mm, %
  basicstyle=\ttfamily\scriptsize,
  keywordstyle=\bfseries\ttfamily,
  commentstyle=\color{darkgray}\itshape\ttfamily,
  keepspaces=true,
  columns=fullflexible,
  escapeinside={(*}{*)},
  mathescape=true,
  showstringspaces=false
}

\lstdefinestyle{inlined} {
  numbers=none,
  frame=none
}

\lstdefinelanguage{JavaRecent}[]{Java}
{
  morekeywords={var,yield}
}

\usepackage{boogie}

\lstdefinelanguage{Grimp}[]{JVMIS}
{
  morekeywords=[2]{int,short,byte,long,char,boolean,float,double,void,class,if,@Require,@Ensure,@Pure,@Predicate,invoke,invokedynamic},
  mathescape=true,
  escapeinside={(*}{*)},
  identifierstyle=\ttfamily,
  keywordstyle={\color{grimpcol}\bfseries\ttfamily},
  keywordstyle=[2]{\color{grimpcol}\bfseries\ttfamily}
}

\lstdefinelanguage{BBlib}[]{JavaRecent}
{
  morekeywords={forall,exists,result,lt,lte,eq,neq,gte,gt,conditional,not,implies,old},
  morekeywords=[2]{invariant,Binding,assertion,assumption},
  morekeywords=[3]{@Require,@Ensure,@Predicate,@Pure},
  deletekeywords=[2]{int,float},
  literate=%
  {:}{$\colon$}1
  {::}{$\bullet$}1
  {!}{$\lnot$}1
  {==}{$=$}1
  {!=}{$\neq$}1
  {&&}{$\land$}1
  {||}{$\lor$}1
  {<}{$<$}1
  {<=}{$\le$}1
  {>}{$>$}1
  {>=}{$\ge$}1
  {==>}{$\Longrightarrow$}3
  {<==>}{$\Longleftrightarrow$}4
  {\\forall}{$\forall$}1
  {\\exists}{$\exists$}1
  ,
  mathescape=true,
  escapeinside={(*}{*)},
  identifierstyle=\ttfamily,
  keywordstyle=[2]{\color{bbcol}\bfseries\ttfamily},
  keywordstyle=[3]{\color{bbcol}\bfseries\ttfamily},
}
 
\newcommand{\J}[1]{\mbox{\lstinline[basicstyle=\ttfamily,language=JavaRecent]|#1|}}

\newcommand{\BBl}[1]{\mbox{\lstinline[basicstyle=\ttfamily,language=BBlib]|#1|}}

\newcommand{\B}[1]{\mbox{\lstinline[basicstyle=\ttfamily,language=boogie,breaklines=true]|#1|}}
\newcommand{\G}[1]{\mbox{\lstinline[basicstyle=\ttfamily,language=Grimp,breaklines=true]|#1|}}
\newcommand{\Tr}{\ensuremath{\mathcal{T}}}
\newcommand{\Agg}{\ensuremath{\mathcal{A}}}
\newcommand{\Fr}{\ensuremath{\mathcal{F}}}
\newcommand{\Ex}{\ensuremath{\mathbb{T}}}

\usepackage{multicol,multirow}

\usepackage{xurl}
\usepackage{fontawesome}

\newcommand{\suppOK}[1][green]{\text{\color{#1}\faCheck}}
\newcommand{\suppNO}[1][red]{\text{\color{#1}\faClose}}
\newcommand{\suppPart}[1][orange]{\text{\color{#1}\faExclamation}}

\newcommand{\byteback}{{\smaller[0.5]{\textsc{Byte\-Back}}}\xspace}
\newcommand{\bblib}{\texttt{BBlib}\xspace}

\let\llncssubparagraph\subparagraph
\let\subparagraph\paragraph
\usepackage[compact]{titlesec}
\let\subparagraph\llncssubparagraph

\iflong
\newcommand{\nicepar}[1]{\paragraph{#1}}
\else
\makeatletter
\newcommand\nicepar{\@startsection{paragraph}{4}{\z@}%
  {-2\p@ \@plus -4\p@ \@minus -4\p@}%
  {-0.5em \@plus -0.22em \@minus -0.1em}%
  {\bfseries\normalsize}}
\makeatother
\fi

\newcommand{\experimentrow}[1]{
  \n[1]{#1/ConversionTime}[0.001]
  & \n[0]{#1/ConversionOverhead}[100]|
  & \n[2]{#1/VerificationTime}[0.001]
  & \n[0]{#1/SourceSize}
  & \n[0]{#1/BytecodeSize}
  & \n[0]{#1/BoogieSize}}

\newcounter{experimentrowcounter}
\newcommand\exprowc{\refstepcounter{experimentrowcounter}\arabic{experimentrowcounter}}

\usepackage{ifthen}

\ifdraft
\usepackage[colorinlistoftodos,textsize=scriptsize,obeyFinal,textwidth=40mm]{todonotes}

\DeclareDocumentCommand{\ReviewNote}{s o m O{white}}{%
  \todo[color=#4,\IfBooleanTF{#1}{inline}{}]{\IfNoValueF{#2}{\textbf{#2:}\xspace}#3}
}
\else
\DeclareDocumentCommand{\ReviewNote}{s o m O{white}}{}
\fi

\DeclareDocumentCommand{\caf}{s m}{\IfBooleanTF{#1}{\ReviewNote*{#2}[yellow]}{\ReviewNote{#2}[yellow]}}
\DeclareDocumentCommand{\mpg}{s m}{\IfBooleanTF{#1}{\ReviewNote*{#2}[red!70!white]}{\ReviewNote{#2}[red!70!white]}}

\begin{document}

\title{Verifying Functional Correctness Properties\\ At the Level of Java Bytecode\ifblind\else\thanks{Work partially supported by SNF grant 200021-207919 (LastMile).}\fi}

\titlerunning{Verifying At the Level of Java Bytecode}

\ifblind
\author{Anonymous authors}
\else
\author{Marco Paganoni\inst{1} \and
Carlo A. Furia\inst{1}\orcidID{0000-0003-1040-3201}}
\authorrunning{M. Paganoni and C. A. Furia}

\institute{Software Institute, USI Università della Svizzera italiana, Lugano, Switzerland\\
\email{marco.paganoni@usi.ch} $\qquad$ \url{bugcounting.net}}
\fi

\maketitle

\SetWatermarkAngle{0}
\SetWatermarkText{\raisebox{14.5cm}{%
   \hspace{0.1cm}%
   \href{https://doi.org/10.5281/zenodo.7337205}{\includegraphics{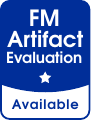}}%
   \hspace{11cm}%
   \includegraphics{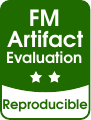}%
}}

\begin{abstract}
  The breakneck evolution of modern programming languages aggravates
  the development of deductive verification tools, which struggle to timely and fully support all new language features.
  To address this challenge, we present \byteback: a verification technique that works on Java bytecode.
  Compared to high-level languages,
  intermediate representations such as bytecode offer a much more limited and stable set of features;
  hence, they may help decouple the verification process from changes in the source-level language.

  \byteback 
  offers a library to specify functional correctness properties at the level of the source code,
  so that the bytecode is only used as an intermediate representation that the end user does not need to work with.
  Then, \byteback reconstructs some of the information about types and expressions
  that is erased during compilation into bytecode but is necessary to correctly perform verification.
  Our experiments with an implementation of \byteback
  demonstrate that it can successfully verify bytecode
  compiled from different versions of Java, and including 
  several modern language features that even state-of-the-art Java verifiers (such as KeY and OpenJML)
  do not directly support---thus
  revealing how \byteback's approach can help keep up verification technology with language evolution.
\end{abstract}

\section{Introduction}
\label{sec:introduction}

Modern programming languages are rich in expressive features and evolve regularly,
extending their capabilities with each new version of the language.
These characteristics make them easier to use and ever more powerful,
to the ultimate benefit of programmers using them.
On the contrary, they also complicate the development of verification tools:
the more features to support, and the faster a programming language evolves,
the harder it is to keep up-to-date a verification toolchain.
Take Java as an example of a widely used modern language.
As we discuss in \autoref{sec:related-work},
no state-of-the-art automated Java verifier fully supports all features of the language%
---even for older versions such as Java~8.

In this paper,
we pursue the idea of performing formal verification not at the level of a language's source code
but on an intermediate representation.
Our \byteback technique processes Java bytecode to verify functional (input/output) properties
expressed as pre- and postconditions. 
By targeting bytecode instead of source code,
\byteback seamlessly supports a wide variety of Java features
that are desugared when automatically translated to bytecode by the compiler.
It can even verify some programs written in other programming languages,
such as Scala, that also compile to Java bytecode. 

Performing functional verification of bytecode entails two main challenges.
First, we need to provide convenient means of expressing the specification to be verified,
as well as any other intermediate annotations. 
Requiring the user to directly annotate the bytecode is impractical,
and at odds with the goal of expressing the behavior of the original Java program.
\byteback offers a Java library (called \bblib) 
with custom annotations and static methods.
Users add specifications to the Java source code by writing Java expressions that call these library methods;
\byteback then recovers the specifications by analyzing \bblib calls in bytecode format.
Supporting expressive contract specifications
of the source code is a key novelty of \byteback compared to other approaches,
such as JayHorn~\cite{JayHorn} and SMACK~\cite{SMACK}, 
that also verify intermediate representations
but mostly target implicit, low-level correctness properties
(such as the absence of memory errors)
and have only limited support for arbitrary functional specifications.

Reconstructing some of the information lost during the compilation from source code to bytecode
is the second main challenge tackled by \byteback.
To this end, we define a bespoke static analysis working on Grimp%
---an alternative representation of bytecode (offered by the Soot static analysis framework~\cite{Soot}).
\byteback's analysis connects bytecode instructions to elements of \bblib specification,
and generates verification conditions that encode program correctness.
Concretely, \byteback translates Grimp code and annotations to the Boogie intermediate verification language~\cite{Boogie},
which we then use as the interface to a backend SMT solver.

We implemented \byteback in a tool with the same name,
which verifies bytecode annotated with functional specifications expressed using the \bblib library.
We verified \exref{ex:last} programs,
including classic verification examples (such as sorting algorithms), 
using numerous Java features that state-of-the-art functional verification tools do not currently support.
We also verified the implementation of some of the same algorithms in Scala,
thus demonstrating that \byteback can accommodate a variety of source-code level features
by focusing on the verification of an intermediate representation.

A replication package including
\byteback's implementation, and 
the benchmarks and examples described in the paper,
is available on Zenodo~\cite{Byteback-REPP}.

\begin{figure}[!hbt]
  \centering
  \begin{subfigure}[t]{0.49\linewidth}
\begin{lstlisting}[language=BBlib]
@Require(forall j: int :: values[j] != 1)
@Ensure(return >= 0)
static int summary1(int[] values) {
	int result = 0;
	for (int k = 0; k < values.length; k++) {
		invariant(result >= 0);
    if (values[k] == 0) result += 1;
    else if (values[k] == 1) result += -1;
    else if (values[k] > 0) result += values[k];
  } return result; }
\end{lstlisting}
    \caption{Method \J{summary1} uses a regular \J{for} loop and \J{if}/\J{else} conditionals.}
    \label{code:counter-old}
  \end{subfigure}
  \hfill
  \begin{subfigure}[t]{0.49\linewidth}
\begin{lstlisting}[language=BBlib]
@Require(forall j: int :: values[j] != 1)
@Ensure(return >= 0)
static int summary2(int... values) {
  var result = 0;
  for (var v: values) result += switch(v) {
  invariant(result >= 0);
    case 0: yield(1);
    case 1: yield(-1);
    default: if (v > 0) yield(v); else yield(0);
  } return result; }
\end{lstlisting}
    \caption{Method \J{summary2} uses varargs,
      a ``foreach'' loop, a \J{switch} expression, and \J{var} local types.}
    \label{code:counter-new}
  \end{subfigure}
  \caption{Annotated Java methods that compute a numeric summary of \J{int} array \J{values}.}
  \label{fig:motivating-examples}
\end{figure}

\section{Motivating Examples}
\label{sec:examples}

\autoref{code:counter-old} shows the implementation of a simple Java method \J{summary1},
which scans its input integer array \J{values} and returns a numeric summary of its content:
it ignores all negative values, adds 1 to the summary for each element 0,
subtracts 1 for each element 1, and adds any bigger positive elements.
The code also embeds some annotations that specify
a precondition \BBl{@Require} (``input array \J{values} includes no element equal to~1''),
a postcondition \BBl{@Ensure} (``the returned result is nonnegative''),
and a loop invariant \BBl{invariant} (``local variable \J{result} stays nonnegative'').
Clearly, \J{summary1} satisfies this specification;
in fact, we can easily verify \autoref{code:counter-old}'s code
against this specification using modern verifiers for Java (such as KeY, Krakatoa, or OpenJML)%
---after expressing the specification using the verifier's annotation language.

Now consider method \J{summary2} in \autoref{code:counter-new}.
It's not hard to see that \J{summary2} implements essentially the same behavior as \J{summary1}
but using different features of the Java language:
\J{values} is a variadic argument (varargs) instead of a plain integer array;
local variable \J{result} uses type inference (\J{var}) instead of declaring its type explicitly;
the loop is an enhanced \J{for} loop (``foreach'');
the loop's body uses a \J{switch} expression with \J{yield} instead of nested \J{if}/\J{else} conditionals.
As shown in \autoref{tab:feature-support}, these features have been added to Java only in recent versions of the language.
As a result, none of the aforementioned Java verifiers that can check the correctness of \J{summary1}
supports all the features used by \J{summary2}%
---even though the methods are essentially equivalent.

Our verification technique \byteback, which we present in the rest of the paper,
performs verification at the level of Java bytecode.
One distinct advantage of this approach is that
the Java compiler takes care of desugaring equivalent Java features into
a simpler representation as bytecode instructions.
Therefore, \byteback verifies both variants \J{summary1} and \J{summary2}
in \autoref{fig:motivating-examples}
without having to explicitly add support for each new Java feature.
This demonstrates that bytecode-level verification can help
formal verification techniques keep up with rapidly evolving source-level languages.

\section{How \byteback Works}
\label{sec:technique}

\begin{figure}[!bt]
\begin{adjustwidth}{-30mm}{-30mm}
\centering
\begin{tikzpicture}[
  bbbox/.style={rectangle,very thick,
    rounded corners=2mm,font=\footnotesize\sffamily,
    minimum width=15mm,minimum height=7mm,
    draw=bbcol,fill=bbcol,text=white,
    label={[below=15pt]#1}},
  toolbox/.style 2 args={rectangle,very thick,
    font=\footnotesize\sffamily,
    draw=black!25,fill=black!25,text=black,
    label={[below=15pt,#2]#1}},  
  databox/.style 2 args={%
    font=\footnotesize,
    label={[below=15pt,#2]#1}},  
  align=center,
  node distance=12mm and 20mm,
  ]

  \matrix[row sep=12mm,column sep=8mm] {
    \node[databox={\texttt{Summary.java}}{name=source-lab}] (source) {source code};
    &
    \node[databox={\texttt{Summary.class}}{name=bytecode-lab},fill=bytecodecol] (bytecode) {bytecode};
    &
    \node[databox={\texttt{Summary.grimp}}{name=grimp-lab},fill=grimpcol,text=white] (grimp) {Grimp IR};
    & 
    \node[databox={\texttt{Summary.bpl}}{name=boogie-code-lab},fill=boogiecol] (boogie-code) {Boogie code};
    \\
    \node[toolbox={\texttt{javac},\ \texttt{scalac},\ \ldots}{name=compiler-lab}] (compiler) {compiler};
    &
    \node[toolbox] (soot) {Soot};
    &
    \node[] (bb-mid) {};
    &
    \node[toolbox] (boogie) {Boogie};
    \\
  };

  \path let \p1 = ($(compiler)!0.5!(soot)$) in
     node[databox={\texttt{bblib.jar}}{name=bblib-lab},fill=bbcol,text=white,
          minimum width=12mm,minimum height=5mm]
          (bblib) at (\x1,\y1-12mm) {\bblib};

  \node[bbbox,above=3pt of bb-mid] (analysis) {analysis};
  \node[bbbox,below=3pt of bb-mid] (encoding) {encoding};
  \node[right=10mm of boogie] (outcome) {};
  \node[above=3pt of outcome] (ok) {\color{green}{\faCheck}};
  \node[below=3pt of outcome] (fail) {\color{red}{\faClose}};

  \node [fit=(analysis)(encoding),draw=bbcol,
         ultra thick,rounded corners,label={[bbcol]south:\textbf{\byteback}}] (byteback) {};

  \begin{scope}[color=black!80,line width=1pt,round cap-latex',every node/.style={font=\footnotesize}]
    \draw (source-lab) -- (compiler);
    \draw (compiler.east) -- ($(compiler)!0.5!(soot)$) |- (bytecode.west);
    \draw (bytecode-lab) -- (soot);
    \draw (soot) -- ($(soot)!0.5!(bb-mid)$) |- (grimp.west);
    \draw (grimp-lab) -- (byteback);
    \draw[color=bbcol] (analysis) -- (encoding);
    \draw[color=bbcol] (byteback.east) -- ($(bb-mid)!0.5!(boogie)$) |- (boogie-code.west);
    \draw (boogie-code-lab) -- (boogie);
    \draw (boogie) -- ($(boogie)!0.6!(outcome)$) -- (ok);
    \draw (boogie) -- ($(boogie)!0.6!(outcome)$) -- (fail);

    \draw (bblib) -| (compiler-lab);
    \draw[-,bbcol] (bblib.east) -- (byteback);
  \end{scope}
  
\end{tikzpicture}
\end{adjustwidth}
\caption{An overview of how \byteback's verification toolchain works.}
\label{fig:workflow}
\end{figure}
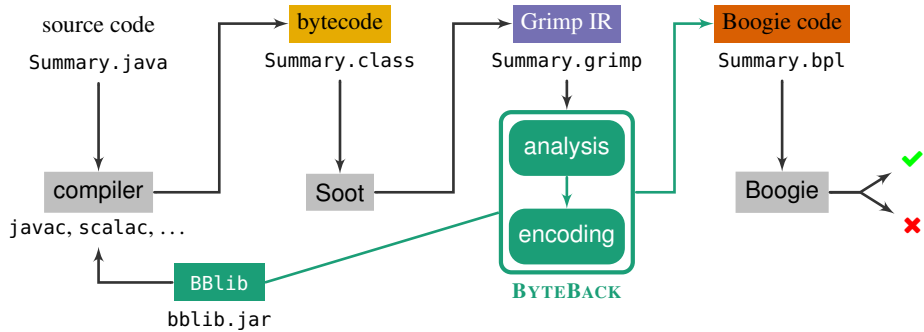

\autoref{fig:workflow} overviews how \byteback works, and the toolchain it implements.
To verify a program,
the user first annotates its \emph{source code} with a specification
using the functionalities of the \bblib library;
\autoref{sec:spec} outlines this library and how it can be used.
\bblib-annotated source code can be compiled
with the Java \emph{compiler} (or any other suitable compiler)
into \emph{bytecode}.
\byteback uses
the Soot static analysis framework
to transform the bytecode into the higher-level \emph{Grimp} intermediate representation
(an alternative bytecode representation
that is syntactically closer to source code and retains higher-level typing information).
As we explain in \autoref{sec:boogie-encoding},
\byteback performs a static \emph{analysis} of Grimp,
with the goal of identifying the various program elements
and linking them to their specification---embedded as calls to \bblib methods, and references to the annotations.
With this information,
\byteback can \emph{translate} program and annotations
into \emph{Boogie} code,
which the Boogie verifier~\cite{Boogie} processes to generate verification conditions,
and finally determine whether the program verifies correctly.

\subsection{Specifying Functional Properties}
\label{subsec:specification}
\label{sec:spec}

This section describes the main methods and annotations
included in the \bblib library,
and how we can use them to express the specification of a Java program.%
\footnote{
\bblib is available as a \textsc{jar} file,
and hence any language that is bytecode-compatible with Java can use its features---as we'll
demonstrate in some of \autoref{sec:experiments}'s examples in Scala.
}
Whereas \autoref{fig:motivating-examples}'s examples use a simplified idiomatic syntax,
in this section we follow \bblib's concrete syntax;
\autoref{code:byteback-summary} shows the same annotations with this concrete syntax.

\begin{figure}[!bth]
\begin{lstlisting}[language=BBlib]
@Require("no_ones")
@Ensure("nonnegative")
public static int summary(int... values);

@Predicate public static boolean no_ones(int[] values)
{ return not(contains(values, 1, 0, values.length)); }

@Predicate public static boolean nonnegative(int[] values, int result)
{ return gte(result, 0); }
	 
@Pure public static boolean contains(int[] as, int e)
{ int i = Binding.integer(); return exists(i, lte(0, i) & lt(i, as.length) & eq(as[i], e)); }
\end{lstlisting}
  \caption{Annotations for \autoref{fig:motivating-examples}'s methods using \bblib's concrete syntax.}
  \label{code:byteback-summary}
\end{figure}

\nicepar{Pre- and postconditions.}
The main specification elements of a method $m$ are its \emph{precondition} and \emph{postcondition},
encoded by adding annotations \BBl{@Require(String p)} and \BBl{@Ensure(String q)}
just before $m$'s declaration.
Arguments \J{p} and \J{q} denote the name of \emph{predicates}:
methods returning \J{boolean} that encode the actual pre- and postconditions.
We can annotate $m$ with several \BBl{@Require}s and \BBl{@Ensure}s,
which are implicitly conjoined.
In \autoref{code:byteback-summary}'s running example,
we name the pre- and postcondition predicates \BBl{no_ones} and \BBl{nonnegative}.

\nicepar{Predicates.}
We mark any predicates $p$ with annotation \BBl{@Predicate}, so that \byteback can easily track them
in the bytecode.
For the same reason, a predicate $p$ is defined in the same class as the method $m$ it specifies.
A predicate $p$ is \J{static} iff $m$ is,
and its input signature types are the same as $m$'s;
this way, $m$'s specification can refer to any program elements that are visible at $m$'s interface.
Since postconditions usually constrain a method's output, 
any predicate $q$ used as a postcondition includes an extra input argument \J{result}
of the same type as $m$'s return type (if it is not \J{void}).
In \autoref{code:byteback-summary}'s running example,
predicates \J{no_ones} and \J{nonnegative} are \J{static} methods like \J{summary};
the latter includes a second argument \BBl{int result},
which refers to the integer value returned by \J{summary}.

\nicepar{Pure expressions.}
A predicate's body
encodes a Boolean expression
that should be exactly expressible in logic.
Therefore, it can only include \emph{pure} (side-effect free)
statements,
and has to terminate with a single \J{return} statement
that defines the overall predicate expression.
In practice, this means
that predicates can only \emph{read} the global program state 
but cannot modify it.
However, pure methods may use local variables
and may call methods
that satisfy the same constraints
and that we marked as \BBl{@Pure}; 
this includes recursive calls.
For example, \autoref{code:byteback-summary}'s
predicate \BBl{no_ones}
calls pure function \BBl{contains}.

\begin{table}[!tbh]
  \centering
  \setlength{\tabcolsep}{4pt}
  \footnotesize
  \begin{tabular}{lcc}
    \toprule
    & \textsc{in Java/logic} & \textsc{in \bblib} \\
    \midrule
    \multirow{2}{*}{comparison}
    & \J{x < y}, \J{x <= y}, \J{x == y} & \BBl{lt(x, y)}, \BBl{lte(x, y)}, \BBl{eq(x, y)}
    \\
    & \J{x != y}, \J{x >= y}, \J{x > y} & \BBl{neq(x, y)}, \BBl{gte(x, y)}, \BBl{gt(x, y)}
    \\[1pt]
    conditionals & \J{c ? t : e} & \BBl{conditional(c, t, e)}
    \\
    \cmidrule(lr){2-3}
    propositional
    & \J{!a}, \J{a && b}, \J{a \|$\!$\| b}, \J{a$\ \Longrightarrow$ b}
    & \BBl{not(a)}, \BBl{a & b}, \BBl{a \| b}, \BBl{implies(a, b)}
    \\[1pt]
    \multirow{2}{*}{quantifiers} & \BBl{$\forall$x: T :: P(x)}& \BBl{T x = Binding.T(); forall(x, P(x))}
    \\
    & \BBl{$\exists$x: T :: P(x)}& \BBl{T x = Binding.T(); exists(x, P(x))}
    \\
    \bottomrule
  \end{tabular}
  \caption{A list of \bblib's aggregable operators, and the
    Java or logic operators that they replace.}
  \label{tab:bblib-ops}
\end{table}

\nicepar{Aggregable expressions.}
\byteback has no access to the source code,
but it should still be able to recover
the pure logic expression encoded by a predicate's body
after this is translated into bytecode by the compiler.
When this is the case, we say that a source code expression is \emph{aggregable}%
---informally, it translates into bytecode without information loss.
Aggregability further constrains what we are allowed to use in a predicate's or pure function's body:
\begin{enumerate*}[label=\emph{\roman*})]
\item Only pure expressions are allowed.
\item Branching statements (conditionals, loops) are not allowed,
  since they introduce jumps in the bytecode that may be cumbersome or impossible to render as a single logic expression.
  Instead, \bblib offers method \BBl{conditional(c, t, e)} to encode conditional \emph{expressions}---similar
  to Java ternary expressions \J{c ? t : e} but translated to bytecode without introducing branching.

\item Java's usual Boolean operators (\J{!}, \J{&&}, \J{||}) are not allowed
  because they are not aggregable:
  \J{&&} and \J{||} are short-circuited, and hence they may introduce branching in the bytecode;
  expressions involving \J{!} may also introduce branching
  (e.g., \J{x = !y} translates to bytecode like \J{if (y) x = false else x = true}).
  Instead, \bblib offers replacement methods (\BBl{not})
  or lets you use Java's eager Boolean operators (\BBl{\&}, \BBl{\|}, \BBl{^}) that are aggregable.

\item Similarly, comparison operators (\J{<}, \J{>}, \ldots)
  may introduce branching in the bytecode, and hence \bblib offers replacement methods
  (\BBl{lt}, \BBl{gt}, \ldots) that are aggregable.
\end{enumerate*}
\autoref{tab:bblib-ops}
summarizes the main aggregable operators provided by \bblib as static methods%
---used either instead of non-aggregable Java methods or to express common logic operators.
\autoref{code:byteback-summary} uses 
some of these operators to express the specification in the running example.

\nicepar{Frame specifications.}
A method's \emph{frame} is the set of memory locations that the method may modify.
\byteback uses a simple approach to \emph{infer} the frame of a method $m$.
It performs a static analysis looking for any heap-modifying statement in $m$'s Boogie translation.%
{}
If it finds any, $m$'s frame is the whole heap; otherwise, $m$'s frame is empty. 
If this analysis determines that $m$'s frame is non-empty but $m$ is marked as \BBl{@Pure} or \BBl{@Predicate},
\byteback reports a verification error.
%
The analysis recursively follows any method called by $m$, and is set up so as to be \emph{sound} but imprecise;
for example, if $m$ calls a method $\ell$ whose implementation is not available,
we conservatively assume that $\ell$ may modify the heap.
Users can still more finely specify a method's frame by adding postconditions
that explicitly indicate heap locations that do \emph{not} change.
Supporting more flexible framing methodologies~\cite{Dynamic,History,Need,Cohen10,DynFrames,SemiCola} in \byteback
belongs to future work.
In \autoref{code:byteback-summary}'s example, 
\byteback infers that \J{summary}'s frame is empty since its implementation only reads the content of array \J{values}.

\nicepar{Other specification elements.}
A method $m$'s postcondition may include 
expressions \BBl{old(e)}---which denotes the value of \BBl{e} in $m$'s pre-state.
In addition,
\bblib offers methods for common intra-method specification elements:
\begin{enumerate*}[label=\emph{\roman*})]
\item \BBl{invariant(J)} declares a loop invariant \BBl{J},
  and can be placed anywhere in the corresponding loop's body.
  \autoref{fig:motivating-examples} shows the loop invariant specification in the running example.

\item methods \BBl{assertion(E)} and \BBl{assumption(E)}
  introduce intermediate assertions (if \J{E} holds continue, otherwise fail)
  and assumptions (ignore states where \J{E} doesn't hold)
  that are useful to further guide the verification process of a method's implementation.
\end{enumerate*}
As usual, the arguments \BBl{J} and \BBl{E} to these specification elements
should be pure, aggregable expressions.

\subsection{Translating Grimp into Boogie}
\label{subsec:conversion}
\label{sec:boogie-encoding}

This section outlines the translation from Grimp%
---a human-readable representation of bytecode produced by the Soot framework---%
to Boogie%
---a verification language that combines an expressive program logic with basic procedural constructs
(variables, assignments, procedures).
Grimp code represents \emph{executable instructions} in a program's bytecode;
in contrast, source-level \emph{declarations} (such as class or variable declarations)
are implicit in Grimp,
but still accessible programmatically through Soot's API.
Concretely, we present \byteback's
Boogie encoding as a translation $\Tr$ from Grimp (instructions) and Java (declarations)
to Boogie code%
---even though this translation
is actually implemented without access to Java source code.
For clarity, we highlight Grimp/Java keywords (\G{goto l})
with a different color than Boogie keywords (\B{goto l}).

\nicepar{Heap model.}
\byteback introduces a simple Boogie model of the heap
adapted from Dafny's~\cite{Dafny}---a state-of-the-art deductive verifier.
The heap is a variable \B{#heap: Heap}\xspace
that stores a polymorphic mapping of \B{type Heap = [Reference]<<\\alpha>>[Field \\alpha]\\alpha}
from references to fields (of generic type \B{\\alpha}).
To access the heap, \byteback defines 
\[
  \B{function read<<\\alpha>>(h:Heap, r:Reference, f:Field \\alpha) returns (\\alpha)}
\]
that returns the value of field \B{f} in the object pointed to by reference \B{r},
and
\[
  \B{function update<<\\alpha>>(h:Heap, r:Reference, f:Field \\alpha, v:\\alpha) returns (Heap)}
\]
that returns an updated heap after setting field \B{r.f} to \B{v}.

\nicepar{Aggregates.}
As we explained in \autoref{sec:spec},
a block of code that defines an \emph{aggregable} expression 
consists of statements that:
\begin{enumerate*}[label=\emph{\roman*})]
\item are \emph{pure} (do not modify the heap);
\item are straight-line (no branches);
\item use \bblib's propositional and comparison operators (\autoref{tab:bblib-ops}),
  or other aggregable user-defined methods.
\end{enumerate*}
Precisely, take a 
sequence $s$ of Grimp instructions
that satisfy these constraints.
Then, $s$ can be written in SSA form~\cite{appel}
as a sequence 
$s_1\,s_2\,\cdots\,s_{n+1}$, $n \geq 0$,
of statements where
each $s_k$, $k \leq n$, is
an assignment $\G{v}_k = e_k$
of an aggregable expression $e_k$ to a fresh variable $\G{v}_k$;
and the final $s_{n+1}$ returns the last $\G{v}_{n}$.
Given any such sequence $s$,
\byteback builds an overall expression $\Agg(s)$
by recursively replacing each usage of $\G{v}_k$ with its unique definition in $s$.
We call $\Agg(s)$ the \emph{aggregate} of snippet $s$;%
\footnote{Soot also performs a kind of aggregation of Grimp expressions; however,
  \byteback's aggregates are different from Soot's in general.
  }
in a nutshell, $\Agg(s)$ is a pure expression equivalent to the one returned by $s$,
which \byteback can translate to a Boogie logic expression
as we detail below.
In \autoref{code:byteback-summary}'s running example,
\BBl{no_ones}'s body is already in aggregate form, and hence
$\Agg(\G{no_ones}) = \G{not(contains(values, 1, 0, values.length))}$.
For convenience, we extend the notation:
$\Agg(e)$, where $e$ is any aggregable expression built by a sequence of statements $s$,
denotes \emph{expression} $e$ in \emph{aggregate form}---defined as $\Agg(s\G{;}\;\G{return}\:e)$.

\nicepar{Types.}
\byteback uses Boogie type \B{int}
(corresponding to mathematical integers)
for all bytecode integer types \G{int}, \G{short}, \G{byte}, \G{long}, and \G{char};
Boogie type \B{real}
(corresponding to mathematical reals)
for floating-point types \G{float} and \G{double};
Boogie type \B{bool}
for type \G{boolean};%
\footnote{While pure bytecode uses \G{0}/\G{1} integers to encode Booleans,
  the Grimp intermediate representation includes a distinct Boolean type \G{boolean}.}
and Boogie type \B{Reference}
for all bytecode reference types.
Thus, for example, $\Tr(\G{int}) = \B{int}$, $\Tr(\G{boolean}) = \B{bool}$, and $\Tr(\G{int[]}) = \B{Reference}$.

\nicepar{Declarations.}
\byteback declares an uninterpreted Boogie type \B{const C: Type} for each \G{class C};
and it declares a \B{const C.f: Field$\;\Tr($t$)$}
for each each field \G{f} of \G{C}---where \G{t} is \G{f}'s static type.%
\footnote{For simplicity, the presentation assumes that identifier names are unique and the same in bytecode as in Boogie;
  in practice, \byteback also takes care of renaming to avoid clashes.}
Similarly, local variables (in implementations of non-pure methods)
translate to Boogie local variables:
$\Tr(\G{t v}) = \B{var v:}\;\Tr(\G{t})$.

\nicepar{Specification functions.}
\byteback translates to Boogie functions any methods annotated with \BBl{@Pure},
which denotes logic functions used in \bblib specifications.
Boogie functions that translate specification functions
include an extra argument \B{h} of type \B{Heap}
since they cannot directly read global variables.
The body $S$ of \BBl{@Pure} methods
has to be aggregable; \byteback first builds the \emph{aggregate} $\Agg(S)$
expression as described above, and then translates that into Boogie.
\autoref{fig:summary-to-boogie} shows the Boogie translation of \J{contains} in the running example.
\begin{align*}
  \Tr\left(
  \begin{array}{l}
    \G{@Pure} \\
    \G{t}_0\ \G{C.p}\;\G{(}\G{t}_1\;\G{d}_1\G{,}\ldots\G{,}\G{t}_m\;\G{d}_m\G{)} \\
    \G{\{}\ S\ \G{\}}
  \end{array}
  \right)
  &=
    \begin{array}{l}
      \B{function C.p} \\
      \quad\;\B{(}\B{h:}\,\B{Heap}\B{,}\B{d}_1\B{:}\,\Tr(\G{t}_1)\B{,}\ldots\B{,}\B{d}_m\B{:}\,\Tr(\G{t}_m)\B{)} \\
      \quad\B{returns}\;\Tr(\G{t}_0)\B{)} \\
      \B{\{}\ \Tr(\Agg(S))\ \B{\}}
    \end{array}
\end{align*}

\begin{figure}[!bt]
\begin{lstlisting}[language=boogie,numbers=none,frame=none,basicstyle=\ttfamily\footnotesize]
procedure summary(values: Reference) returns (@ret: int)
requires !contains(#heap, values, 1, 0, lengthof(values)); ensures @ret >= 0;

function contains(h: Heap, as: Reference, e: int) returns(bool)
{ exists i: int :: 0 <= i && i <= lengthof(as) && (array.read(h, as, i): int) == e }
\end{lstlisting}
  \caption{\byteback's Boogie encoding of \J{summary}'s signature and \J{contains} in \autoref{code:byteback-summary}.}
  \label{fig:summary-to-boogie}
\end{figure}

\nicepar{Methods.}
\byteback translates to Boogie procedures any other methods (that is, not annotated with \BBl{@Pure} or \BBl{@Predicate}).
An additional extra argument \B{o} of type \B{Reference}
matches the \emph{target} of method calls; thus, it is absent in procedures translating static methods.
Methods that return a value (whose return type is not \G{void})
include a return argument named \B{@ret} in Boogie, which is also passed to the postcondition predicate.
Frame specifications 
translate to Boogie \B{modifies} clauses;
\byteback infers them as described in \autoref{sec:spec},
and hence they can only be empty (the \B{modifies} clause is omitted)
or include the whole heap (\B{modifies \#heap}).
Preconditions and postconditions translate to Boogie \B{requires} and \B{ensures} clauses
as follows.
Given a \BBl{@Predicate} method \G{p},
\byteback first builds its aggregate expression $\Agg(\G{p})$;
then, it translates this Grimp expression to a Boogie expression $\Tr(\Agg(\G{p}))$;
finally, it replaces \G{p}'s formal arguments with the corresponding Boogie formal arguments $\B{d}_1,\ldots,\B{d}_m$.
\begin{align*}
  \Tr\left(
  \begin{array}{l}
    \G{@Require("p")} \\
    \G{@Ensure("q")} \\
    \G{t}_0\ \G{C.m}\;\G{(}\G{t}_1\;\G{d}_1\G{,}\ldots\G{,}\G{t}_m\;\G{d}_m\G{)} \\
    \G{\{}\ B\ \G{\}}
  \end{array}
  \right)
  &=
    \begin{array}{l}
      \B{procedure C.m}\\
      \;\B{(}\B{o}\B{:}\,\B{Reference}\B{,}\,\B{d}_1\B{:}\,\Tr(\G{t}_1)\B{,}\ldots\B{,}\B{d}_m\B{:}\,\Tr(\G{t}_m)\B{)} \\
      \quad\B{returns (@ret:}\,\Tr(\G{t}_0)\B{)} \\
      \quad\B{requires}\ \Tr(\Agg(\G{p}))[\B{d}_1\B{,}\ldots\B{,}\B{d}_m]\\
      \quad\B{ensures}\ \Tr(\Agg(\G{q}))[\B{d}_1\B{,}\ldots\B{,}\B{d}_m\B{,}\B{@ret}]\\
      \quad\B{modifies}\ \Fr(B)\\
      \B{\{}\ \Tr(B)\ \B{\}}
    \end{array}
\end{align*}
\autoref{fig:summary-to-boogie} shows the Boogie translation of
\J{summary}'s signature and specification.
Why does \byteback translate postconditions in this way (inlining aggregate specification expressions),
instead of just using the Boogie functions that translate postcondition predicates%
---such as \B{nonnegative(values, @result)} for \J{summary}'s postcondition?
In general, postconditions may use \G{old} to refer to an expression's value in the pre-state;
Boogie offers an \B{old} operator, but only accepts it explicitly in an \B{ensures}, %
not in user-defined functions.
Therefore, a postcondition \G{@Ensure("inc")}, where predicate \G{inc} is declared as
\G{@Predicate boolean inc()\{return gt(x, old(x));\}}
can only be translated as
\B{ensures read(\#heap,this,C.x) > old(read(\#heap,this,C.x))}%
---not as \B{ensures inc(#heap)}, since \B{inc}'s body may not use \B{old}.

\emph{Constructors}
may also have a specification.
\byteback translates them like special methods that return a fresh (previously unallocated)
reference in the heap to the created object---as specified by an automatically generated postcondition.
To this end, \byteback supplies Boogie procedures \B{new} and \B{array.new} to create new references,
which translate bytecode instructions \G{new} and \G{newarray}.
Then, actual constructor calls (\G{invokespecial} in bytecode) translate like
normal procedure calls---as shown below.

\nicepar{Expected types.}
Expression types in Grimp 
mirror strictly the bytecode instructions they correspond to.
This may lead to Soot attributing to a Grimp expression $e$
an unnecessarily general type $t$ when $e$ is actually only used according
to a more specific type $t'$.
For example, the type of Grimp expression $a\ \G{&}\ b$ is \G{int} according to
Soot even if $a$ and $b$ are of type \G{boolean}.
To have more specific types in these scenarios,
\byteback reconstructs the \emph{expected type} $\Ex(e)$
of any Grimp expression $e$ based on where $e$ is used.
Thus, if $e$ is the right-hand side of an assignment $\G{v} = e$,
$\Ex(e)$ is \G{v}'s type;
if $e$ is returned by a method $m$, 
$\Ex(e)$ is $m$'s return type according to its signature;
if $e$ is the actual argument in a call to $m$,
$\Ex(e)$ is $m$'s formal argument type.
Therefore, $\Ex(a\ \G{&}\ b)$ is \G{boolean} as long as $a\ \G{&}\ b$ is used as a Boolean.





\begin{table}[!htb]
  \centering
  \begin{tabular}{lll}
    \toprule
    \textsc{Grimp: $e$} & \textsc{Boogie: $\Tr(e)$} & \\
    \midrule
    $\G{v}$ & $\B{v}$ & local variable read \\
    $\G{o.f}$ & $\B{read(\#heap, o, f)}$ & instance field read \\
    $\G{C.f}$ & $\B{read(\#heap, type2ref(C), f)}$ & static field read \\
    $\G{a[}k\G{]}$ & $\B{array.read(\#heap, a,}\:\Tr(k)\B{):}\:\Tr(\Ex(\G{a[}k\G{]}))$ & array read \\
    \midrule
    $\G{v}\ \G{=}\ e$ & $\B{v :=}\ \Tr(e)$ & local variable write \\
    $\G{o.f}\ \G{=}\ e$ & $\B{\#heap := update(\#heap, o, f,}\:\Tr(e)\B{)}$ & instance field write \\
    $\G{C.f}\ \G{=}\ e$ & $\B{\#heap := update(\#heap, type2ref(C), f,}\,\Tr(e)\B{)}$ & static field write \\
    $\G{a[}k\G{]}\ \G{=}\ e$ & $\B{\#heap := array.update(\#heap, a,}\:\Tr(k)\B{,}\,\Tr(e)\B{)}$ & array write \\
    \bottomrule
  \end{tabular}
  \caption{Boogie translation of read and write of variables in Grimp bytecode.}
  \label{tab:read-write-vars}
\end{table}

\nicepar{Variable access.}
\autoref{tab:read-write-vars} summarizes
the translation of reading and writing variables (local, instance, static, and array).
Local variables are straightforward, as they also are local variables in Boogie.
\emph{Fields} of objects in the heap are read and written by calling
predefined Boogie functions \B{read} and \B{heap.write} introduced earlier in this section.
\emph{Unqualified} field accesses \G{f} translate 
as qualified accesses \G{this.f} on \G{this}---which corresponds to some
variable of type \B{Reference} in Boogie.
The same functions \B{read} and \B{heap.write} also work for \emph{static} field accesses:
to this end, \byteback supplies 
\[
  \B{function type2ref(class: Type) returns(Reference)}
\]
mapping each class type to a reference to a heap object that stores the static state.
\emph{Arrays} are also heap objects, but \byteback provides custom functions \B{array.read} and \B{array.update}
to access these objects by means of an index expression of type \B{int}.
\[
  \B{function array.read<<\\alpha>>(h: Heap, a: Reference, k: int) returns (\\alpha)}
\]
As shown in \autoref{tab:read-write-vars},
\byteback casts (Boogie `\B{:}' operator)
the output of polymorphic \B{array.read} to array type $\Ex(\G{a[}k\G{]})$.
This is not necessary for field accesses,
since \B{read}'s output type parameter \B{\\alpha} is constrained by the input \B{f};
in contrast, \B{array.read}
is only type-generic in the output,
and hence usage context determines the concrete value of \B{\\alpha}.

\nicepar{Calls.}
Bytecode offers five call \emph{instructions}:
\G{invokestatic} (to call \J{static} methods),
\G{invokevirtual} (instance methods),
\G{invokeinterface} (abstract interface calls),
\G{invokespecial} (constructors and \J{super} calls),
and, since Java~7, \G{invokedynamic} (lambdas).
\byteback translates all such call \emph{instructions} to Boogie procedure calls:%
\footnote{Thus, \byteback relies on Boogie's \emph{modular} semantics of calls:
  the only effects of calling a method \B{m} are what \B{m}'s specification prescribes.
  This is a standard choice in deductive verification, since it supports modularity and
  is consistent with the Liskov substitution principle~\cite{meyer-axiomaticsemantics}.
  }
\[
  \Tr(\G{invokevirtual o.m(}e_1, \ldots, e_n\G{)}) =
  \B{call C.m(o,}\:\Tr(e_1)\B{,}\ldots\B{,}\Tr(e_n)\B{)}
\]
As usual, \B{C} is \B{m}'s class, and \B{o} is a reference to an instance of this class.
The same translation works, with obvious adjustments, for
the other kinds of call instructions%
---except \G{invokedynamic}, which \byteback doesn't currently support.
Henceforth, \G{invoke} denotes any of the four supported bytecode call instructions.

\nicepar{Branching.}
\byteback translates branching instructions (\G{return}, \G{goto}, and \G{if})
into the corresponding Boogie statements
as shown in \autoref{tab:branching}.
While Boogie also offers structured conditionals and loops,
\byteback does not use them since 
bytecode does not have structured programming constructs.

\nicepar{Literals.}
\byteback translates any literal $\ell$ to a Boogie literal according to its expected type $\Ex(\ell)$.
In particular, $\Tr(\G{0}) = \B{false}$ and $\Tr(\G{1}) = \B{true}$
when the expected type of integer literals \G{0} and \G{1} is \G{boolean}.

\begin{table}
  \setlength{\tabcolsep}{3pt}
  \centering
  \begin{subtable}{0.7\textwidth}
    \centering
    \begin{tabular}[t]{lll}
      \toprule
      \textsc{Grimp: $e$} & \textsc{Boogie: $\Tr(e)$} & 
      \\
      \midrule
      $\G{return}\ v$ & $\B{@ret :=}\ \Tr(v)\B{; return}$
      \\
      $\G{goto}\ \ell$ & $\B{goto}\ \ell$
      \\
      $\G{if (}c\G{)}\:B$ & $\B{if (}\Tr(c)\B{) \{}\Tr(B)\B{\}}$
      \\
      \bottomrule
    \end{tabular}
    \caption{Boogie encoding of Grimp bytecode branching instructions.}
    \label{tab:branching}
  \end{subtable}
  \\
  \begin{subtable}{0.4\textwidth}
    \centering
    \begin{tabular}{lll}
      \toprule
      \textsc{Grimp: $e$} & \textsc{Boogie: $\Tr(e)$} & 
      \\
      \midrule
      $\G{neg}\;a$ & $\B{!}\Tr(a)$ & not
      \\
      $a\;\G{^}\;b$ & $\Tr(a)\;\B{!=}\;\Tr(b)$ & xor
      \\
      $a\;\G{&}\;b$ & $\Tr(a)\;\B{&&}\;\Tr(b)$ & and
      \\
      $a\;\G{|}\;b$ & $\Tr(a)\;\B{||}\;\Tr(b)$ & or
      \\
      $\G{implies(}a\G{,}b\G{)}$ & $\Tr(a)\;\B{==>}\;\Tr(b)$ & implies
      \\
      $a\;\G{==}\;b$ & $\Tr(a)\;\B{<==>}\;\Tr(b)$ & iff
      \\
      \bottomrule
    \end{tabular}
    \caption{Boogie encoding of Grimp bytecode Boolean operators. Grimp expressions $a$ and $b$ have expected type $\G{boolean} = \Ex(a) = \Ex(b)$.}
    \label{tab:boolean-operators}
  \end{subtable}
  \hspace{7mm}
  \begin{subtable}{0.45\textwidth}
    \centering
    \begin{tabular}{ll}
      \toprule
      \textsc{Grimp: $e$} & \textsc{Boogie: $\Tr(e)$}
      \\
      \midrule
      $\G{assertion(}b\G{)}$ & $\B{assert}\ \Tr(\Agg(b))$
      \\
      $\G{assumption(}b\G{)}$ & $\B{assume}\ \Tr(\Agg(b))$
      \\
      $\G{forall(v,}\:b\G{)}$ & $\B{forall v:}\ \Tr(\Ex(\G{v}))\;\B{::}\;\Tr(\Agg(b))$
      \\
      $\G{exists(v,}\:b\G{)}$ & $\B{exists v:}\ \Tr(\Ex(\G{v}))\;\B{::}\;\Tr(\Agg(b))$
      \\
      $\begin{array}{l}
         \G{conditional}\\
         \quad\G{(}b\G{,}T\G{,}E\G{)}
       \end{array}$
      & $\begin{array}{l}
           \B{if}\;\Tr(\Agg(b))\;\\
           \quad\B{then}\;\Tr(\Agg(T))\;\\
           \quad\B{else}\;\Tr(\Agg(E))
         \end{array}$
      \\
      \bottomrule
    \end{tabular}
    \caption{Boogie encoding of \bblib specification methods and expressions. Variables \B{v} are created by methods of class \BBl{Binding}.}
    \label{tab:trans-spec-operators}
  \end{subtable}
  \caption{\byteback translation of branching, Boolean operators and specification elements.}
  \label{tab:operators-branching}
\end{table}

\nicepar{Expressions.}
Most arithmetic and comparison operators
$\G{+},\,\G{-},\, \G{*},\, \G{==},\, \G{!=},\, \G{<},\, \G{<=},\, \G{>=},\, \G{>}$
translate to their
Boogie counterparts
\ $\B{+},\, \B{-},\, \B{*},\, \B{==},\, \B{!=},\, \B{<},\, \B{<=},\, \B{>=},\, \B{>}$
as obvious: $\Tr(a \bowtie b) = \Tr(a)\ \Tr(\bowtie)\ \Tr(b)$.
The division operator \G{/} translates to \B{div} or \B{/} in Boogie according to whether it represent integer or floating-point division:
$\Tr(a\;\G{/}\;b) = \Tr(a)\;\B{div}\;\Tr(b)$ if $\Ex(a\;\G{/}\;b) = \G{int}$;
otherwise $\Tr(a\;\G{/}\;b) = \Tr(a)\;\B{/}\;\Tr(b)$.
\byteback introduces and axiomatizes an overloaded Boogie function \B{cmp}
to translate bytecode operator \G{cmp}: $\Tr(a\;\G{cmp}\;b) = \B{cmp(}\Tr(a)\B{,}\Tr(b)\B{)}$
returns \G{1} if $a > b$, \G{-1} if $a < b$, and \G{0}
if $a = b$.
\autoref{tab:boolean-operators}
displays how \byteback translates Grimp Boolean operators to Boogie.
Java's short-circuited operators \J{&&} and \J{||} are not listed in the table,
as the compiler desugars them into \emph{conditional instructions} in bytecode;
for example, \J{if (a && b) x = 1 $\ldots$} becomes \G{if (a == 0) goto end; if (b == 0) goto end; x = 1; end:$\ldots$}
in bytecode.

Since \G{boolean} is a subtype of \G{int} in Soot,
the operands of Boolean operator expressions
(e.g., $a\:\G{==}\:b$)
may have different types (e.g., $\Ex(a) = \G{int}$ but $\Ex(b) = \G{boolean}$%
---usually when $a$ is used as an integer in other parts of the program).
In these cases, \byteback translates everything using the most general type \B{int},
so that all usages of the operands can be uniformly represented in Boogie
(where the \B{bool} and \B{int} types are disjoint, as they are in Java).

\nicepar{Call expressions.}
Boogie does not allow procedure calls in expressions;%
\footnote{In contrast, calls to pure methods, translated to Boogie functions,
  can be directly transliterated to Boogie (pure) expressions.
}
therefore, \byteback saves the call value in a fresh variable,
and replaces the call expression with a read of the variable:
given a Grimp expression $e$, used in statement $s$, 
that includes a call \G{invoke o.m()} (virtual, static, or interface) to a method \G{m},
\byteback first adds the statements 
$\B{var \#r:}\:\Tr(\Ex(\G{invoke o.m}))\B{; call \#r :=}\ \Tr(\G{invoke o.m()})$
just before $s$, 
and then translates $e$ into $\Tr(e[\G{invoke o.m} \mapsto \B{\#r}])$---replacing the call with \B{\#r}.

\nicepar{Specifications.}
\byteback recognize \bblib operators and translates them to their counterparts in Boogie,
as shown in \autoref{tab:trans-spec-operators}.
Source-code \J{while} and \J{for} loops
become conditional jumps in bytecode.
Using Soot's static analysis capabilities,
\byteback identifies any loop in bytecode
by its \emph{head}, \emph{backjump}, and \emph{exit} locations.
Thus, a source-code loop $\J{while (!c) }\;L\J{;}\ R$
corresponds to bytecode structured as in \autoref{tab:loop-encoding}'s
left-hand side,
where labels \underline{\G{head}}, \underline{\G{back}}, and \underline{\G{exit}}
mark the head, exit, and backjump locations;
$c$ is the loop's exit condition, $B$ is the loop body,
and $R$ is the code that follows the loop.
Any loop invariant $J$ 
would be declared by a call $\G{invariant(}J\G{)}$
to \bblib method \BBl{invariant}
inside $B$.
\byteback encodes the semantics of loop invariants by
means of suitable assumptions and assertions,
as in \autoref{tab:loop-encoding}'s right-hand side;
then,
it translates the annotated branching code to Boogie as usual.
\autoref{tab:aggregation-in-loop-inv} shows a concrete example of how \byteback encodes loops and invariants; note the aggregation (inlining) of the invariant predicate, which ensures that all its dependencies are replicated in each assertion and assumption in Boogie.

\begin{figure}[!hbt]
  \begin{subfigure}{1.0\linewidth}
  \setlength{\tabcolsep}{8pt}
  \centering
  \begin{tabular}{ll}
    ${\underline{\G{head:}}}\ \G{if (}c\G{) goto exit}$ & $\G{assert(}\Agg(J)\G{);}\ \underline{\G{head:}}\ \G{if (}c\G{) goto exit}$ \\
    $B[\cdots\:\G{invariant(}J\G{)} \cdots]$ & $\G{assume(}\Agg(J)\G{);}\ B[\cdots]$ \\
    ${\underline{\G{back:}}}\ \G{goto head}$ & $\G{assert(}\Agg(J)\G{);}\ {\underline{\G{back:}}}\ \G{goto head}$ \\
    ${\underline{\G{exit:}}}\ R$ & ${\underline{\G{exit:}}}\ \G{assume(}\Agg(J)\G{);}\ R$
  \end{tabular}
  \caption{How \byteback injects loop invariant checks (right) into Grimp bytecode loops (left).
    $B$ denotes the loop body, which includes an invariant declaration (left).}
  \label{tab:loop-encoding}
\end{subfigure}
  \begin{subfigure}{1.0\linewidth}
  \setlength{\tabcolsep}{8pt}
  \centering
  \begin{tabular}{ll}
    $\BBl{for (int k = 0; k < 3; k++)}$
    & $\B{assert(0 <= k && k <= 3);}$ \\
    $\BBl{\{}\quad \BBl{boolean a = lte(0, k);}$
    & $\underline{\B{head}:}\:\B{if (k >= 3) goto exit;}$ \\
    $\ \;\quad \BBl{boolean b = lte(k, 3);}$
    & $\B{assume(0 <= k && k <= 3); k := k + 1;}$ \\
    $\ \;\quad \BBl{invariant(a & b); \}}$
                        & $\B{assert(0 <= k && k <= 3);}\ \underline{\B{back:}}\:\B{goto head;}$ \\
    & $\underline{\B{exit:}}\:\B{assume(0 <= k && k <= 3);}$
  \end{tabular}
  \caption{An example of an annotated loop in Java (left), and its Boogie encoding produced by \byteback (right).}
  \label{tab:aggregation-in-loop-inv}
\end{subfigure}
\caption{\byteback's encoding of loops and loop invariants.}
  \label{tab:loop-encoding-general}
\end{figure}

\subsection{Implementation Details}
\label{sec:implementation}

We implemented \byteback in a tool with the same name,
written in about 11 thousand lines of Java code (plus another 52 kLOC of generated code).
\byteback's core uses the Soot static analysis framework~\cite{Soot} to process the bytecode to be verified,
as we described in \autoref{subsec:conversion} at a high level.
After analyzing the Grimp bytecode,
\byteback has collected all the information to generate Boogie code;
to this end, a visitor pattern implementation creates a Boogie AST,
and then dumps it into a Boogie file.

We developed the Boogie AST library using the metacompilation framework Jast\-Add~\cite{JastAdd}, 
in combination with JFlex and Beaver%
\footnote{JFlex: \url{https://jflex.de/}; Beaver: \url{http://beaver.sourceforge.net/}.}
to parse Boogie source code.
This capability is useful to:
\begin{enumerate*}[label=\emph{\roman*})]
\item flexibly generate the heap model (\autoref{subsec:conversion}) and other Boogie background definitions
  from a human-readable Boogie template file;
\item perform some analyses directly on the generated Boogie code
  (most notably, the frame inference briefly described in \autoref{subsec:specification}).
\end{enumerate*}

\nicepar{Features and limitations.}
\byteback's current implementation supports
most bytecode features
but not exception handling and \G{invokedynamic} (which limits reasoning about lambdas);
strings are supported as plain objects, which precludes precisely analyzing their semantics in Java;
as we discussed previously, numeric types are encoded as infinite-precision numbers (integers and reals),
which entails that \byteback may miss overflow and other numerical errors.
Adding support for these features is possible in principle,
and would require a combination of extending the Boogie encoding
(for example, to include exceptional behavior),
\byteback's static analysis
(for example, to identify the bootstrap methods that dynamically generate targets of \G{invokedynamic}),
and \bblib's features
(for example, to support model-based specifications).


\autoref{subsec:specification} described the features offered by \byteback's \bblib specification library.
Its current implementation is sufficient to specify a variety of examples (see \autoref{sec:experiments})
but lacks advanced features to express complex framing conditions and
ghost code (specification code discarded during compilation),
and to flexibly reuse specifications with inheritance and modularity.
Supporting these features belongs to future work,
also because it would require tackling challenges largely orthogonal to the focus of \byteback.

As we demonstrate in \autoref{sec:experiments},
\bblib's features can also specify programs written in Scala,
leveraging its bytecode-level interoperability with Java.
However, \bblib was developed with focus on Java,
and hence its practical usability on Scala is more limited.
In particular, the Scala compiler automatically generates
features (such as setters and getters for fields)
that are implicit in Scala source code;
hence, users cannot directly annotate these features using \bblib.
Addressing these limitations is possible,
but would have to cater somewhat 
to the peculiarities of Scala (or other languages to be supported).

\section{Experiments}
\label{sec:experiments}

In our experiments, we ran \byteback on several examples in order to demonstrate
that it can verify programs with different characteristics,
which exercise various features of the Java programming language (including recent versions),
as well as a few programs written in other languages built on top of Java bytecode.

\subsection{Programs}
\label{sec:exp-subjects}

\autoref{tab:experiments} lists the \exref{ex:first}[ex:last]
programs that we used for our experiments,
and their size in non-empty lines of \textsc{source} code (LOC),
as well as their size after compilation to \textsc{bytecode} (also in LOC of the representation returned by \lstinline[language=bash,basicstyle=\ttfamily\footnotesize]{javap -c}).
The sizes include the annotations in \bblib, which specify the properties to be verified.

The majority of programs
(\exref{ex:j8-first}[ex:j8-last]/\exref{ex:first}[ex:last])
use various features of Java~8;
but we also included \exref{ex:j17-first}[ex:j17-last]
programs using Java~17 features,
and \exref{ex:s-first}[ex:s-last]
Scala programs.
The selection includes relatively simple programs
that specifically target language features of Java (examples \exref{ex:j8-first-simple}--\exref{ex:j8-last-simple} and \exref{ex:j17-first-simple}--\exref{ex:j17-last-simple}),
classic algorithms and procedures
(examples \exref{ex:j8-first-proc}--\exref{ex:j8-last-proc}, \exref{ex:j17-first-proc},
and \exref{ex:s-first-proc}--\exref{ex:s-last-proc}),
and object-oriented features
(examples \exref{ex:j8-first-oo}--\exref{ex:j8-last-oo}
and \exref{ex:s-first-oo}--\exref{ex:s-last-oo}).
Some examples implement the same algorithm 
for data structures with different types (e.g., \J{double} and \J{int} arrays).

We selected these examples to demonstrate that \byteback
can process a variety of modern Java features,
including several that state-of-the-art Java deductive verifiers
do not support
(as we discuss in \autoref{sec:feature-support}).
It's important to stress that we are not comparing \byteback's verification capabilities
to those of much more mature tools such as KeY, Krakatoa, and OpenJML.
We picked the features in \autoref{tab:experiments}'s examples 
specifically to demonstrate that it's hard for source-level verifiers
to keep up with the plethora of language features that are introduced over time%
---not to solve verification challenges.
As we discuss in \autoref{sec:technique},
\byteback does not support all used features of Java
(in particular, exceptions)
and its specification capabilities (in particular, framing)
are currently limited compared to source-level tools.
The experiments only demonstrate our
claim that verification at the level of bytecode has some distinctive advantages
for supporting language evolution, and hence it can complement
source-level verification.

\begin{table}[!tb]
  \setlength{\tabcolsep}{2pt}
  \centering
  \scriptsize
  \begin{tabular}{rllrrrrrr}
    \toprule
    \multicolumn{1}{c}{\textsc{\#}}
             & \multicolumn{1}{c}{\textsc{experiment}}
             & \multicolumn{1}{c}{\textsc{language}}
             & \multicolumn{2}{c}{\textsc{encoding}}
             & \multicolumn{1}{c}{\textsc{verification}}
             & \multicolumn{1}{c}{\textsc{source}}
             & \multicolumn{1}{c}{\textsc{bytecode}}
             & \multicolumn{1}{c}{\textsc{Boogie}}
                                                                                                   \\
             & 
             & 
             & \multicolumn{1}{c}{\textsc{time} [s]}
             & \multicolumn{1}{c}{\byteback}
             & \multicolumn{1}{c}{\textsc{time} [s]}
             & \multicolumn{3}{c}{\textsc{size} [LOC]}
                                                                                                   \\
    \midrule
    \exprowc\label{ex:j8-first-simple}\label{ex:j8-first}\label{ex:first}
             & Array Operations          & Java 8 & \experimentrow{/j8/array/Basic}               \\
    \exprowc
             & Boolean Operations        & Java 8 & \experimentrow{/j8/operation/Boolean}          \\
    \exprowc
             & Control Flow              & Java 8 & \experimentrow{/j8/controlflow/Basic}         \\
    \exprowc\label{ex:loopinfer1}
             & Enhanced For              & Java 8 \tiny{F} & \experimentrow{/j8/controlflow/EnhancedFor}    \\
    \exprowc
             & Field Access              & Java 8 & \experimentrow{/j8/instance/FieldAccess}       \\
    \exprowc
             & Floating-Point Operations & Java 8 & \experimentrow{/j8/operation/Real}             \\
    \exprowc
             & Instance Field            & Java 8 & \experimentrow{/j8/instance/InstanceField}     \\
    \exprowc
             & Integer Operations        & Java 8 & \experimentrow{/j8/operation/Integer}             \\
    \exprowc
             & Multiclass                & Java 8 & \experimentrow{/j8/instance/Supported}         \\
    \exprowc
             & Quantifiers               & Java 8 & \experimentrow{/j8/quantifier/Basic}          \\
    \exprowc
             & Static Field              & Java 8 & \experimentrow{/j8/instance/StaticField}       \\
    \exprowc
             & Static Initializer        & Java 8 & \experimentrow{/j8/instance/StaticInitializer} \\
    \exprowc
             & Static Method Calls       & Java 8 & \experimentrow{/j8/operation/StaticCall}       \\
    \exprowc
             & Switch                    & Java 8 & \experimentrow{/j8/controlflow/Switch}         \\
    \exprowc
             & Unit                      & Java 8 & \experimentrow{/j8/instance/Unit}              \\
    \exprowc\label{ex:j8-last-simple}
             & Virtual Method Calls      & Java 8 & \experimentrow{/j8/operation/VirtualCall}      \\

    \exprowc\label{ex:j8-first-proc}
             & GCD                       & Java 8 & \experimentrow{/j8/algorithm/GCD}                  \\
    \exprowc
             & Insertion Sort \J{double} & Java 8 & \experimentrow{/j8/algorithm/DoubleInsertionSort}  \\
    \exprowc
             & Insertion Sort \J{int}    & Java 8 & \experimentrow{/j8/algorithm/IntegerInsertionSort} \\
    \exprowc
             & Linear Search             & Java 8 \tiny{T} & \experimentrow{/j8/algorithm/LinearSearch}         \\
    \exprowc
             & Max \J{double}            & Java 8 & \experimentrow{/j8/algorithm/DoubleMax}           \\
    \exprowc
             & Max \J{int}               & Java 8 & \experimentrow{/j8/algorithm/IntegerMax}           \\
    \exprowc
             & Selection Sort \J{double} & Java 8 & \experimentrow{/j8/algorithm/DoubleSelectionSort}  \\
    \exprowc
             & Selection Sort \J{int}    & Java 8 & \experimentrow{/j8/algorithm/IntegerSelectionSort} \\
    \exprowc
             & Square Sorted Array       & Java 8 & \experimentrow{/j8/algorithm/SquareSortedArray}    \\
    \exprowc
             & Sum \J{double}            & Java 8 & \experimentrow{/j8/algorithm/DoubleSum}            \\
    \exprowc\label{ex:j8-last-proc}
             & Sum \J{int}               & Java 8 & \experimentrow{/j8/algorithm/IntegerSum}           \\

    \exprowc\label{ex:j8-first-oo}
             & Generic List       & Java 8 \tiny{G, T}  & \experimentrow{/j8/generics/List}                 \\
    \exprowc
             & Binary Search      & Java 8  & \experimentrow{/j8/algorithm/BinarySearch}        \\
    \exprowc
             & Comparator         & Java 8  & \experimentrow{/j8/instance/Comparator}           \\
    \exprowc
             & Dice               & Java 8 \tiny{D}  & \experimentrow{/j8/instance/Dice}                 \\
    \exprowc\label{ex:j8-last-oo}\label{ex:j8-last}
             & Counter            & Java 8  & \experimentrow{/j8/instance/Counter}              \\

    \exprowc\label{ex:j17-first-simple}\label{ex:j17-first}
             & Pattern matching   & Java 17 \tiny{P} & \experimentrow{/j17/patternmatching/Basic}       \\
    \exprowc
             & Switch Expressions & Java 17 \tiny{S, Y} & \experimentrow{/j17/switchexpression/Basic} \\
    \exprowc\label{ex:j17-last-simple}
             & Type Inference     & Java 17 \tiny{L} & \experimentrow{/j17/typeinference/Basic}         \\

    \exprowc\label{ex:j17-first-proc}\label{ex:j17-last-proc}\label{ex:j17-last}\label{ex:loopinfer2}
             & Summary                   & Java 17 \tiny{S, Y, F, L, A}  & \experimentrow{/j17/algorithm/Summary}             \\

    \exprowc\label{ex:s-first-proc}\label{ex:s-first}
             & GCD                       & Scala 3 & \experimentrow{/s/complete/GCD}                    \\
    \exprowc\label{ex:s-last-proc}
             & Linear Search             & Scala 3 & \experimentrow{/s/complete/LinearSearch}           \\

    \exprowc\label{ex:s-first-oo}
             & Comparator                & Scala 3 & \experimentrow{/s/complete/Comparator}             \\
    \exprowc\label{ex:s-last-oo}\label{ex:s-last}\label{ex:last}
             & Dice                      & Scala 3 & \experimentrow{/s/complete/Dice}                   \\
    \midrule
    & \textbf{total}                     &
    & \n[1]{/total/ConversionTime}[0.001]
    &
    & \n[2]{/total/VerificationTime}[0.001]
    & \n[0]{/total/SourceSize}
    & \n[0]{/total/BytecodeSize}
    & \n[0]{/total/BoogieSize}\\
    & \textbf{average}                   &
    & \n[1]{/average/ConversionTime}[0.001]
    & \n[0]{/average/ConversionOverhead}[100]|
    & \n[2]{/average/VerificationTime}[0.001]
    & \n[0]{/average/SourceSize}
    & \n[0]{/average/BytecodeSize}
    & \n[0]{/average/BoogieSize}
    \\
    \bottomrule
  \end{tabular}
  \caption{Verification experiments performed with \byteback. %
    In each row: the \textsc{experiment}'s name;
    its source \textsc{language} (and any of \autoref{tab:feature-support}'s features it uses);
    the \textsc{encoding time} (seconds)
    and its percentage directly attributable to \byteback 
    (excluding Soot's initialization time);
    the \textsc{verification time} (seconds)
    of running Boogie on the encoding generated by \byteback;
    the size (in non-blank lines of code)
    of the \textsc{source} code,
    of the \textsc{bytecode}
    (as printed by \lstinline[language=bash,basicstyle=\ttfamily\footnotesize]{javap -c}),
    and of the \textsc{Boogie} code.
  }
  \label{tab:experiments}
\end{table}

\subsection{Results}
\label{sec:exp-results}

All the experiments ran \byteback
on a Fedora~36 GNU/Linux machine with an Intel i7-7600U CPU (2.8GHz), 
running Boogie~2.15.7.0, Z3~4.11.1.0, and Soot~4.3.0.
To account for possible measurement noise,
we repeated each experiment 5 times 
and report the mean of the wall-clock running times in the 95th percentile.

We ran Boogie with default options%
---except for programs~\exref{ex:loopinfer1} and~\exref{ex:loopinfer2},
where we enabled option \texttt{/infer:j},
which can infer simple 
loop invariants.
This is useful to handle these programs' enhanced \J{for} loops:
translated to bytecode,
a loop such as \J{for(var v: values)} in \autoref{code:counter-new}
introduces an index variable \J{int k} to iterate over array \J{values};
however, \J{k} does not exist in the source code,
and hence one cannot annotate the loop with a suitable invariant for \J{k}
and must rely on inferring it.

All the experiments in \autoref{tab:experiments}
verified successfully without errors.
The running time of \byteback (column \textsc{encoding time})
is generally short and predictable:
\n[1]{/average/ConversionTime}[0.001] seconds per example on average.
This time measures \byteback's analysis of bytecode
and translation to Boogie;
it excludes the compilation time (from Java/Scala to bytecode)
and the running time of Boogie (reported separately in column \textsc{verification time}).
Column \textsc{encoding byteback} reports
the percentage of encoding time
after we deduct Soot's fixed context initialization time:
\byteback's net average analysis time is
a small fraction of the total
(just \n[2]{/average/ConversionTime}[0.00009] seconds per example).

The running time of Boogie (column \textsc{verification time})
on \byteback's output
is also moderate: \n[1]{/average/VerificationTime}[0.001]
seconds per example on average.
There are a few outliers: the two variants of Selection Sort take
up to \n[0]{/j8/algorithm/DoubleSelectionSort/VerificationTime}[0.001]
seconds to verify.
This is because Selection Sort's implementation
calls another method to compute the minimum value in an array range;
this introduces more modular verification work.
In contrast, Insertion Sort's implementation uses two nested loops,
which results in a simpler Boogie program.

If we compare \autoref{tab:experiments}'s two rightmost columns,
we notice that the size of the Boogie code
is roughly proportional to the size of the bytecode
(Kendall's $\tau = \n[2]{/stat/cor(bytecode size, Boogie size)}$).
Boogie code is about  
\n[1]{/stat/Boogie conciseness}
times larger,
as \byteback's aggregation process reconstructs complex higher-level expressions.
The size difference is especially pronounced
for programs focusing on object-oriented features
(examples \exref{ex:j8-first-oo}--\exref{ex:j8-last-oo}
and \exref{ex:s-first-oo}--\exref{ex:s-last-oo}):
such features are desugared in bytecode,
but ``resurface'' in the form of Boogie axioms and functions.

These experiments demonstrate \byteback's current capabilities.
Its Boogie encoding is fairly standard
(as mentioned in \autoref{subsec:conversion}, its heap model is taken from Dafny's)
but could be optimized for better performance
(e.g., improving triggers~\cite{triggers-why,stabilizing-triggers,CF-iFM17})
or for conciseness
(e.g., further simplifying type conversions~\cite{Whiley})
as needed.

\section{Related Work}
\label{sec:related-work}

We summarize related work in the areas most relevant to \byteback:
source-level deductive verifiers for Java, and verifiers that target intermediate representations. 

\iflong
\else
  \nicepar{Source-level deductive verifiers for Java.}
\fi
  \label{sec:feature-support}
Performing deductive verification of functional properties on a program's source code
is a widespread approach, as that's where a specification and other kinds of information
are readily available and naturally expressible.
Among the many source-level verifiers for realistic programming languages%
---e.g., \cite{VeriFast,FramaC,VCC,SpecSharp,VerCors,ScalaStainless,Dafny,AutoProof}---%
here we focus on KeY~\cite{KeY,KeYbook2}, Krakatoa~\cite{Krakatoa}, and OpenJML~\cite{OpenJML}:
state-of-the-art verifiers for the functional correctness of Java sequential programs
with a high degree of automation.

OpenJML and Krakatoa follow the so-called auto-active approach~\cite{AutoActive}%
---where
the verifier generates verification conditions (VCs) and dispatches them to an automated theorem prover,
but the user still indirectly guides the verifier by interactively supplying annotations.
OpenJML generates VCs in SMT-LIB format~\cite{SMTLIB},
and dispatches them to any SMT solver like Z3~\cite{Z3} or CVC4~\cite{CVC4}.
Krakatoa translates the source program into the WhyML intermediate verification language IVL, 
and delegates the generation of VCs to the Why3 system~\cite{Why3}.
Using an IVL to generate VCs is an approach pioneered by Spec\#~\cite{SpecSharp}
and used nowadays by many systems (including \byteback).
KeY is built on top of an interactive prover for Java dynamic logic~\cite{DynLogic}%
---used as its intermediate representation---%
but offers features that increase the automation level in practice.

\begin{table}[!tb]
  \centering
  \scriptsize
  \input{tables/feature-data.tex}
\caption{%
    Features of the Java language, and which source-code verifiers support them.
    For each \textsc{feature}: the Java major \textsc{version} when it was introduced,
    an \textsc{example} snippet of code using the feature,
    and which Java verifier among Key, Krakatoa, and OpenJML
    supports (\suppOK), partially supports (\suppPart), or does not support (\suppNO) the feature.
  }
  \label{tab:feature-support}
\end{table}

KeY, Krakatoa, and OpenJML all use JML~\cite{JML} as specification language
(more precisely, different variants/subsets of JML~\cite{JML-differences}).
Despite being applicable to verify real-world Java code,
they also differ in the subset of Java that they support:
\autoref{tab:feature-support} lists several modern features of the Java language
and which verifier can analyze them.
We compiled the table by reading the tools' official documentation and papers,
and by trying out the latest tool versions that are publicly available.
It should be clear that this summary is not a criticism of KeY, Krakatoa, or OpenJML%
---which are state-of-the-art, mature tools
with proven applicability to complex verification problems---%
nor a direct comparison with \byteback.
To compile \autoref{tab:feature-support},
we actively looked for Java recent feature ``variants''
that may be cumbersome to support at the source-code level,
but are essentially syntactic sugar.
Since \byteback easily supports these features
by piggybacking off the compiler's bytecode translation,
this substantiates our claim that
keeping verification tools up to pace with language evolution is practically hard
and time-consuming at the source-code level, but substantially easier at the bytecode level.

The difference in feature support reflects the tools' intended verification target.
Krakatoa focuses on supporting complex functional specifications of a core subset of Java;
thus, it ignores several features that have been available since Java~5
(released in 2002).
KeY and OpenJML aim at verifying complex, realistic Java applications~\cite{KeY-TimSort-proof,KeY-LinkedList-proof,KeY-IdentityHashMap-proof,OpenJML-in-practice};
to this end, they enjoy a broader language support and at least parse all Java features up until version~8
(released in 2014); 
however, several widely used features are still not available for verification with these tools.
For example, KeY relies on an external tool to erase generics and replace them with type \J{Object}
and suitable casts;
OpenJML natively supports generics but not all related features---such as the diamond operator \J{<>}.
Since Java switched to a biannual release schedule,
the gap between available language features and verification support
has been widening~\cite{OpenJML-java16}.

\iflong
\else
  \nicepar{Verifiers for intermediate representations.}
\fi
%
Approaches targeting the verification of intermediate representations (IRs)
have been introduced in recent years,
including SeaHorn~\cite{SeaHorn} and SMACK~\cite{SMACK} for LLVM bitcode~\cite{LLVM}, 
and JayHorn~\cite{JayHorn} for Java bytecode.
A key difference between \byteback and these tools are
the kinds of properties they are equipped to verify:
SMACK, SeaHorn, and JayHorn mainly target low-level implicit correctness properties
(such as the absence of unreachable code, null pointer dereferences, and out-of-bound accesses);
users can still add simple inline assertions, but there is no support
for complex and structured specification elements such as contracts.
SeaHorn and JayHorn encode IR instructions into constrained Horn clauses~\cite{CHC}, 
a logic that can be automatically analyzed with symbolic model-checking techniques.
This is consistent with these tools' intended usage,
as it requires fewer annotations (loop invariants can often be inferred automatically)
but also somewhat restricts the properties that can be verified in practice.
SMACK, like \byteback, translates an IR into Boogie programs to perform verification;
despite these similarities, it mainly targets the verification
of low-level (e.g., embedded) programs~\cite{SMACK-CANAL,SMACK-SoC,SMACK-opt-comp}
and properties;
it defaults to bounded verification (full, unbounded verification is only experimentally supported).

\emph{Proof-carrying code}~\cite{pcc} is another application
of verification techniques to IRs.
To ensure a safe execution,
compiled programs are distributed with embedded proofs, 
which the runtime environment checks before starting execution.
Due to the difficulty of verifying IRs, proof-carrying code was primarily
used for restricted properties such as memory safety.
\emph{Proof-transformation} approaches~\cite{proof-transf,gb-pcc} overcome this issue 
by first verifying source-level annotated program ``as usual'',
and then transforming the correctness proofs into proof-carrying IR code~\cite{B2BPL}.
The BML notation takes a different approach~\cite{BML-tools}
to directly annotate bytecode with expressive JML-like specifications.

\section{Conclusions}
\label{sec:conclusions}

We presented \byteback, a technique that formally verifies functional source-code properties
by working on Java bytecode.
In our experiments, we verified programs written in Java that use recently introduced features
that even state-of-the-art verifiers do not fully support; as well as some programs written in Scala
that \byteback can also analyze after compiling to bytecode.
This suggests that our approach can help simplify keeping up with the evolution of modern programming languages,
which regularly add new expressive features that are substantially simplified by compilation to bytecode.

\bibliographystyle{splncs04}

\input{byteback.bbl}
\end{document}

%% file: experiments.tex
\pgfkeyssetvalue{/bb/j8/algorithm/BinarySearch/ConversionTime}{3074.6}
\pgfkeyssetvalue{/bb/j8/algorithm/BinarySearch/ConversionOverhead}{0.0982015907873012}
\pgfkeyssetvalue{/bb/j8/algorithm/BinarySearch/VerificationTime}{1133.6}
\pgfkeyssetvalue{/bb/j8/algorithm/BinarySearch/SourceSize}{51}
\pgfkeyssetvalue{/bb/j8/algorithm/BinarySearch/BytecodeSize}{124}
\pgfkeyssetvalue{/bb/j8/algorithm/BinarySearch/BoogieSize}{131}
\pgfkeyssetvalue{/bb/j8/algorithm/DoubleInsertionSort/ConversionTime}{2935.6}
\pgfkeyssetvalue{/bb/j8/algorithm/DoubleInsertionSort/ConversionOverhead}{0.1060466098265568}
\pgfkeyssetvalue{/bb/j8/algorithm/DoubleInsertionSort/VerificationTime}{2526.4}
\pgfkeyssetvalue{/bb/j8/algorithm/DoubleInsertionSort/SourceSize}{49}
\pgfkeyssetvalue{/bb/j8/algorithm/DoubleInsertionSort/BytecodeSize}{132}
\pgfkeyssetvalue{/bb/j8/algorithm/DoubleInsertionSort/BoogieSize}{147}
\pgfkeyssetvalue{/bb/j8/algorithm/DoubleMax/ConversionTime}{3019.6}
\pgfkeyssetvalue{/bb/j8/algorithm/DoubleMax/ConversionOverhead}{0.0907256456281415}
\pgfkeyssetvalue{/bb/j8/algorithm/DoubleMax/VerificationTime}{1153.4}
\pgfkeyssetvalue{/bb/j8/algorithm/DoubleMax/SourceSize}{45}
\pgfkeyssetvalue{/bb/j8/algorithm/DoubleMax/BytecodeSize}{92}
\pgfkeyssetvalue{/bb/j8/algorithm/DoubleMax/BoogieSize}{92}
\pgfkeyssetvalue{/bb/j8/algorithm/DoubleSelectionSort/ConversionTime}{3107.6}
\pgfkeyssetvalue{/bb/j8/algorithm/DoubleSelectionSort/ConversionOverhead}{0.1263410786323203}
\pgfkeyssetvalue{/bb/j8/algorithm/DoubleSelectionSort/VerificationTime}{4562.8}
\pgfkeyssetvalue{/bb/j8/algorithm/DoubleSelectionSort/SourceSize}{87}
\pgfkeyssetvalue{/bb/j8/algorithm/DoubleSelectionSort/BytecodeSize}{231}
\pgfkeyssetvalue{/bb/j8/algorithm/DoubleSelectionSort/BoogieSize}{172}
\pgfkeyssetvalue{/bb/j8/algorithm/DoubleSum/ConversionTime}{3015.0}
\pgfkeyssetvalue{/bb/j8/algorithm/DoubleSum/ConversionOverhead}{0.0796627402067477}
\pgfkeyssetvalue{/bb/j8/algorithm/DoubleSum/VerificationTime}{1164.4}
\pgfkeyssetvalue{/bb/j8/algorithm/DoubleSum/SourceSize}{35}
\pgfkeyssetvalue{/bb/j8/algorithm/DoubleSum/BytecodeSize}{70}
\pgfkeyssetvalue{/bb/j8/algorithm/DoubleSum/BoogieSize}{124}
\pgfkeyssetvalue{/bb/j8/algorithm/GCD/ConversionTime}{3011.8}
\pgfkeyssetvalue{/bb/j8/algorithm/GCD/ConversionOverhead}{0.0852897719395999}
\pgfkeyssetvalue{/bb/j8/algorithm/GCD/VerificationTime}{1145.4}
\pgfkeyssetvalue{/bb/j8/algorithm/GCD/SourceSize}{41}
\pgfkeyssetvalue{/bb/j8/algorithm/GCD/BytecodeSize}{88}
\pgfkeyssetvalue{/bb/j8/algorithm/GCD/BoogieSize}{127}
\pgfkeyssetvalue{/bb/j8/algorithm/IntegerInsertionSort/ConversionTime}{3081.4}
\pgfkeyssetvalue{/bb/j8/algorithm/IntegerInsertionSort/ConversionOverhead}{0.1061397160692086}
\pgfkeyssetvalue{/bb/j8/algorithm/IntegerInsertionSort/VerificationTime}{1704.4}
\pgfkeyssetvalue{/bb/j8/algorithm/IntegerInsertionSort/SourceSize}{49}
\pgfkeyssetvalue{/bb/j8/algorithm/IntegerInsertionSort/BytecodeSize}{131}
\pgfkeyssetvalue{/bb/j8/algorithm/IntegerInsertionSort/BoogieSize}{147}
\pgfkeyssetvalue{/bb/j8/algorithm/IntegerMax/ConversionTime}{2926.2}
\pgfkeyssetvalue{/bb/j8/algorithm/IntegerMax/ConversionOverhead}{0.0986391855967819}
\pgfkeyssetvalue{/bb/j8/algorithm/IntegerMax/VerificationTime}{1158.4}
\pgfkeyssetvalue{/bb/j8/algorithm/IntegerMax/SourceSize}{45}
\pgfkeyssetvalue{/bb/j8/algorithm/IntegerMax/BytecodeSize}{90}
\pgfkeyssetvalue{/bb/j8/algorithm/IntegerMax/BoogieSize}{126}
\pgfkeyssetvalue{/bb/j8/algorithm/IntegerSelectionSort/ConversionTime}{3034.2}
\pgfkeyssetvalue{/bb/j8/algorithm/IntegerSelectionSort/ConversionOverhead}{0.1202066702525069}
\pgfkeyssetvalue{/bb/j8/algorithm/IntegerSelectionSort/VerificationTime}{3561.2}
\pgfkeyssetvalue{/bb/j8/algorithm/IntegerSelectionSort/SourceSize}{87}
\pgfkeyssetvalue{/bb/j8/algorithm/IntegerSelectionSort/BytecodeSize}{230}
\pgfkeyssetvalue{/bb/j8/algorithm/IntegerSelectionSort/BoogieSize}{172}
\pgfkeyssetvalue{/bb/j8/algorithm/IntegerSum/ConversionTime}{2801.2}
\pgfkeyssetvalue{/bb/j8/algorithm/IntegerSum/ConversionOverhead}{0.0797497412874262}
\pgfkeyssetvalue{/bb/j8/algorithm/IntegerSum/VerificationTime}{1144.2}
\pgfkeyssetvalue{/bb/j8/algorithm/IntegerSum/SourceSize}{35}
\pgfkeyssetvalue{/bb/j8/algorithm/IntegerSum/BytecodeSize}{70}
\pgfkeyssetvalue{/bb/j8/algorithm/IntegerSum/BoogieSize}{124}
\pgfkeyssetvalue{/bb/j8/algorithm/LinearSearch/ConversionTime}{2876.4}
\pgfkeyssetvalue{/bb/j8/algorithm/LinearSearch/ConversionOverhead}{0.1017531187430792}
\pgfkeyssetvalue{/bb/j8/algorithm/LinearSearch/VerificationTime}{1140.2}
\pgfkeyssetvalue{/bb/j8/algorithm/LinearSearch/SourceSize}{60}
\pgfkeyssetvalue{/bb/j8/algorithm/LinearSearch/BytecodeSize}{126}
\pgfkeyssetvalue{/bb/j8/algorithm/LinearSearch/BoogieSize}{164}
\pgfkeyssetvalue{/bb/j8/algorithm/SquareSortedArray/ConversionTime}{2913.8}
\pgfkeyssetvalue{/bb/j8/algorithm/SquareSortedArray/ConversionOverhead}{0.0961083367607012}
\pgfkeyssetvalue{/bb/j8/algorithm/SquareSortedArray/VerificationTime}{1136.8}
\pgfkeyssetvalue{/bb/j8/algorithm/SquareSortedArray/SourceSize}{54}
\pgfkeyssetvalue{/bb/j8/algorithm/SquareSortedArray/BytecodeSize}{123}
\pgfkeyssetvalue{/bb/j8/algorithm/SquareSortedArray/BoogieSize}{140}
\pgfkeyssetvalue{/bb/j8/array/Basic/ConversionTime}{2772.4}
\pgfkeyssetvalue{/bb/j8/array/Basic/ConversionOverhead}{0.0953178076805433}
\pgfkeyssetvalue{/bb/j8/array/Basic/VerificationTime}{1162.4}
\pgfkeyssetvalue{/bb/j8/array/Basic/SourceSize}{36}
\pgfkeyssetvalue{/bb/j8/array/Basic/BytecodeSize}{103}
\pgfkeyssetvalue{/bb/j8/array/Basic/BoogieSize}{148}
\pgfkeyssetvalue{/bb/j8/casting/BoolToInt/ConversionTime}{2695.4}
\pgfkeyssetvalue{/bb/j8/casting/BoolToInt/ConversionOverhead}{0.0673020213703321}
\pgfkeyssetvalue{/bb/j8/casting/BoolToInt/VerificationTime}{1158.4}
\pgfkeyssetvalue{/bb/j8/casting/BoolToInt/SourceSize}{20}
\pgfkeyssetvalue{/bb/j8/casting/BoolToInt/BytecodeSize}{23}
\pgfkeyssetvalue{/bb/j8/casting/BoolToInt/BoogieSize}{107}
\pgfkeyssetvalue{/bb/j8/casting/IntToReal/ConversionTime}{2852.8}
\pgfkeyssetvalue{/bb/j8/casting/IntToReal/ConversionOverhead}{0.0651256461217509}
\pgfkeyssetvalue{/bb/j8/casting/IntToReal/VerificationTime}{1135.2}
\pgfkeyssetvalue{/bb/j8/casting/IntToReal/SourceSize}{27}
\pgfkeyssetvalue{/bb/j8/casting/IntToReal/BytecodeSize}{45}
\pgfkeyssetvalue{/bb/j8/casting/IntToReal/BoogieSize}{107}
\pgfkeyssetvalue{/bb/j8/casting/RealToInt/ConversionTime}{2845.8}
\pgfkeyssetvalue{/bb/j8/casting/RealToInt/ConversionOverhead}{0.0589803517668304}
\pgfkeyssetvalue{/bb/j8/casting/RealToInt/VerificationTime}{1107.6}
\pgfkeyssetvalue{/bb/j8/casting/RealToInt/SourceSize}{17}
\pgfkeyssetvalue{/bb/j8/casting/RealToInt/BytecodeSize}{22}
\pgfkeyssetvalue{/bb/j8/casting/RealToInt/BoogieSize}{97}
\pgfkeyssetvalue{/bb/j8/controlflow/Basic/ConversionTime}{2793.2}
\pgfkeyssetvalue{/bb/j8/controlflow/Basic/ConversionOverhead}{0.0838666714805794}
\pgfkeyssetvalue{/bb/j8/controlflow/Basic/VerificationTime}{1329.6}
\pgfkeyssetvalue{/bb/j8/controlflow/Basic/SourceSize}{74}
\pgfkeyssetvalue{/bb/j8/controlflow/Basic/BytecodeSize}{123}
\pgfkeyssetvalue{/bb/j8/controlflow/Basic/BoogieSize}{219}
\pgfkeyssetvalue{/bb/j8/controlflow/EnhancedFor/ConversionTime}{2866.8}
\pgfkeyssetvalue{/bb/j8/controlflow/EnhancedFor/ConversionOverhead}{0.0835420741361955}
\pgfkeyssetvalue{/bb/j8/controlflow/EnhancedFor/VerificationTime}{1251.6}
\pgfkeyssetvalue{/bb/j8/controlflow/EnhancedFor/SourceSize}{25}
\pgfkeyssetvalue{/bb/j8/controlflow/EnhancedFor/BytecodeSize}{52}
\pgfkeyssetvalue{/bb/j8/controlflow/EnhancedFor/BoogieSize}{107}
\pgfkeyssetvalue{/bb/j8/controlflow/Switch/ConversionTime}{3072.8}
\pgfkeyssetvalue{/bb/j8/controlflow/Switch/ConversionOverhead}{0.0644795530931034}
\pgfkeyssetvalue{/bb/j8/controlflow/Switch/VerificationTime}{1233.6}
\pgfkeyssetvalue{/bb/j8/controlflow/Switch/SourceSize}{23}
\pgfkeyssetvalue{/bb/j8/controlflow/Switch/BytecodeSize}{25}
\pgfkeyssetvalue{/bb/j8/controlflow/Switch/BoogieSize}{109}
\pgfkeyssetvalue{/bb/j8/generics/List/ConversionTime}{3069.8}
\pgfkeyssetvalue{/bb/j8/generics/List/ConversionOverhead}{0.0887744426615895}
\pgfkeyssetvalue{/bb/j8/generics/List/VerificationTime}{1170.6}
\pgfkeyssetvalue{/bb/j8/generics/List/SourceSize}{46}
\pgfkeyssetvalue{/bb/j8/generics/List/BytecodeSize}{68}
\pgfkeyssetvalue{/bb/j8/generics/List/BoogieSize}{134}
\pgfkeyssetvalue{/bb/j8/instance/Comparator/ConversionTime}{3003.8}
\pgfkeyssetvalue{/bb/j8/instance/Comparator/ConversionOverhead}{0.1048853204356813}
\pgfkeyssetvalue{/bb/j8/instance/Comparator/VerificationTime}{1241.8}
\pgfkeyssetvalue{/bb/j8/instance/Comparator/SourceSize}{51}
\pgfkeyssetvalue{/bb/j8/instance/Comparator/BytecodeSize}{30}
\pgfkeyssetvalue{/bb/j8/instance/Comparator/BoogieSize}{188}
\pgfkeyssetvalue{/bb/j8/instance/Counter/ConversionTime}{2935.2}
\pgfkeyssetvalue{/bb/j8/instance/Counter/ConversionOverhead}{0.0817277857171208}
\pgfkeyssetvalue{/bb/j8/instance/Counter/VerificationTime}{1204.0}
\pgfkeyssetvalue{/bb/j8/instance/Counter/SourceSize}{33}
\pgfkeyssetvalue{/bb/j8/instance/Counter/BytecodeSize}{62}
\pgfkeyssetvalue{/bb/j8/instance/Counter/BoogieSize}{150}
\pgfkeyssetvalue{/bb/j8/instance/Dice/ConversionTime}{2999.4}
\pgfkeyssetvalue{/bb/j8/instance/Dice/ConversionOverhead}{0.0958756328433227}
\pgfkeyssetvalue{/bb/j8/instance/Dice/VerificationTime}{1170.4}
\pgfkeyssetvalue{/bb/j8/instance/Dice/SourceSize}{41}
\pgfkeyssetvalue{/bb/j8/instance/Dice/BytecodeSize}{25}
\pgfkeyssetvalue{/bb/j8/instance/Dice/BoogieSize}{129}
\pgfkeyssetvalue{/bb/j8/instance/FieldAccess/ConversionTime}{2837.0}
\pgfkeyssetvalue{/bb/j8/instance/FieldAccess/ConversionOverhead}{0.0648290376713456}
\pgfkeyssetvalue{/bb/j8/instance/FieldAccess/VerificationTime}{1177.2}
\pgfkeyssetvalue{/bb/j8/instance/FieldAccess/SourceSize}{29}
\pgfkeyssetvalue{/bb/j8/instance/FieldAccess/BytecodeSize}{32}
\pgfkeyssetvalue{/bb/j8/instance/FieldAccess/BoogieSize}{96}
\pgfkeyssetvalue{/bb/j8/instance/InstanceField/ConversionTime}{2950.6}
\pgfkeyssetvalue{/bb/j8/instance/InstanceField/ConversionOverhead}{0.0570246571063068}
\pgfkeyssetvalue{/bb/j8/instance/InstanceField/VerificationTime}{1151.2}
\pgfkeyssetvalue{/bb/j8/instance/InstanceField/SourceSize}{18}
\pgfkeyssetvalue{/bb/j8/instance/InstanceField/BytecodeSize}{16}
\pgfkeyssetvalue{/bb/j8/instance/InstanceField/BoogieSize}{98}
\pgfkeyssetvalue{/bb/j8/instance/StaticField/ConversionTime}{4414.4}
\pgfkeyssetvalue{/bb/j8/instance/StaticField/ConversionOverhead}{0.0916428597249557}
\pgfkeyssetvalue{/bb/j8/instance/StaticField/VerificationTime}{1819.0}
\pgfkeyssetvalue{/bb/j8/instance/StaticField/SourceSize}{32}
\pgfkeyssetvalue{/bb/j8/instance/StaticField/BytecodeSize}{66}
\pgfkeyssetvalue{/bb/j8/instance/StaticField/BoogieSize}{146}
\pgfkeyssetvalue{/bb/j8/instance/StaticInitializer/ConversionTime}{3724.2}
\pgfkeyssetvalue{/bb/j8/instance/StaticInitializer/ConversionOverhead}{0.063597664627277}
\pgfkeyssetvalue{/bb/j8/instance/StaticInitializer/VerificationTime}{1633.8}
\pgfkeyssetvalue{/bb/j8/instance/StaticInitializer/SourceSize}{14}
\pgfkeyssetvalue{/bb/j8/instance/StaticInitializer/BytecodeSize}{14}
\pgfkeyssetvalue{/bb/j8/instance/StaticInitializer/BoogieSize}{91}
\pgfkeyssetvalue{/bb/j8/instance/Supported/ConversionTime}{3044.6}
\pgfkeyssetvalue{/bb/j8/instance/Supported/ConversionOverhead}{0.0803988494015586}
\pgfkeyssetvalue{/bb/j8/instance/Supported/VerificationTime}{1190.6}
\pgfkeyssetvalue{/bb/j8/instance/Supported/SourceSize}{14}
\pgfkeyssetvalue{/bb/j8/instance/Supported/BytecodeSize}{14}
\pgfkeyssetvalue{/bb/j8/instance/Supported/BoogieSize}{113}
\pgfkeyssetvalue{/bb/j8/instance/Unit/ConversionTime}{2921.0}
\pgfkeyssetvalue{/bb/j8/instance/Unit/ConversionOverhead}{0.0575190950047567}
\pgfkeyssetvalue{/bb/j8/instance/Unit/VerificationTime}{1151.0}
\pgfkeyssetvalue{/bb/j8/instance/Unit/SourceSize}{13}
\pgfkeyssetvalue{/bb/j8/instance/Unit/BytecodeSize}{12}
\pgfkeyssetvalue{/bb/j8/instance/Unit/BoogieSize}{97}
\pgfkeyssetvalue{/bb/j8/operation/Boolean/ConversionTime}{3596.6}
\pgfkeyssetvalue{/bb/j8/operation/Boolean/ConversionOverhead}{0.0896584812368424}
\pgfkeyssetvalue{/bb/j8/operation/Boolean/VerificationTime}{1339.6}
\pgfkeyssetvalue{/bb/j8/operation/Boolean/SourceSize}{57}
\pgfkeyssetvalue{/bb/j8/operation/Boolean/BytecodeSize}{85}
\pgfkeyssetvalue{/bb/j8/operation/Boolean/BoogieSize}{157}
\pgfkeyssetvalue{/bb/j8/operation/Integer/ConversionTime}{3070.6}
\pgfkeyssetvalue{/bb/j8/operation/Integer/ConversionOverhead}{0.1203506460826936}
\pgfkeyssetvalue{/bb/j8/operation/Integer/VerificationTime}{1448.6}
\pgfkeyssetvalue{/bb/j8/operation/Integer/SourceSize}{202}
\pgfkeyssetvalue{/bb/j8/operation/Integer/BytecodeSize}{332}
\pgfkeyssetvalue{/bb/j8/operation/Integer/BoogieSize}{250}
\pgfkeyssetvalue{/bb/j8/operation/Real/ConversionTime}{2928.4}
\pgfkeyssetvalue{/bb/j8/operation/Real/ConversionOverhead}{0.0783586941841013}
\pgfkeyssetvalue{/bb/j8/operation/Real/VerificationTime}{1213.2}
\pgfkeyssetvalue{/bb/j8/operation/Real/SourceSize}{37}
\pgfkeyssetvalue{/bb/j8/operation/Real/BytecodeSize}{52}
\pgfkeyssetvalue{/bb/j8/operation/Real/BoogieSize}{110}
\pgfkeyssetvalue{/bb/j8/operation/StaticCall/ConversionTime}{3007.8}
\pgfkeyssetvalue{/bb/j8/operation/StaticCall/ConversionOverhead}{0.0757257179941079}
\pgfkeyssetvalue{/bb/j8/operation/StaticCall/VerificationTime}{1185.6}
\pgfkeyssetvalue{/bb/j8/operation/StaticCall/SourceSize}{32}
\pgfkeyssetvalue{/bb/j8/operation/StaticCall/BytecodeSize}{40}
\pgfkeyssetvalue{/bb/j8/operation/StaticCall/BoogieSize}{112}
\pgfkeyssetvalue{/bb/j8/operation/VirtualCall/ConversionTime}{2959.0}
\pgfkeyssetvalue{/bb/j8/operation/VirtualCall/ConversionOverhead}{0.0691108397690959}
\pgfkeyssetvalue{/bb/j8/operation/VirtualCall/VerificationTime}{1210.4}
\pgfkeyssetvalue{/bb/j8/operation/VirtualCall/SourceSize}{31}
\pgfkeyssetvalue{/bb/j8/operation/VirtualCall/BytecodeSize}{40}
\pgfkeyssetvalue{/bb/j8/operation/VirtualCall/BoogieSize}{122}
\pgfkeyssetvalue{/bb/j8/quantifier/Basic/ConversionTime}{3117.4}
\pgfkeyssetvalue{/bb/j8/quantifier/Basic/ConversionOverhead}{0.0625845765041526}
\pgfkeyssetvalue{/bb/j8/quantifier/Basic/VerificationTime}{1152.2}
\pgfkeyssetvalue{/bb/j8/quantifier/Basic/SourceSize}{25}
\pgfkeyssetvalue{/bb/j8/quantifier/Basic/BytecodeSize}{28}
\pgfkeyssetvalue{/bb/j8/quantifier/Basic/BoogieSize}{92}
\pgfkeyssetvalue{/bb/j17/algorithm/Summary/ConversionTime}{3073.4}
\pgfkeyssetvalue{/bb/j17/algorithm/Summary/ConversionOverhead}{0.0966654854361789}
\pgfkeyssetvalue{/bb/j17/algorithm/Summary/VerificationTime}{1230.8}
\pgfkeyssetvalue{/bb/j17/algorithm/Summary/SourceSize}{47}
\pgfkeyssetvalue{/bb/j17/algorithm/Summary/BytecodeSize}{88}
\pgfkeyssetvalue{/bb/j17/algorithm/Summary/BoogieSize}{137}
\pgfkeyssetvalue{/bb/j17/patternmatching/Basic/ConversionTime}{2981.4}
\pgfkeyssetvalue{/bb/j17/patternmatching/Basic/ConversionOverhead}{0.0663107713504266}
\pgfkeyssetvalue{/bb/j17/patternmatching/Basic/VerificationTime}{1136.8}
\pgfkeyssetvalue{/bb/j17/patternmatching/Basic/SourceSize}{18}
\pgfkeyssetvalue{/bb/j17/patternmatching/Basic/BytecodeSize}{26}
\pgfkeyssetvalue{/bb/j17/patternmatching/Basic/BoogieSize}{105}
\pgfkeyssetvalue{/bb/j17/switchexpression/Basic/ConversionTime}{2865.4}
\pgfkeyssetvalue{/bb/j17/switchexpression/Basic/ConversionOverhead}{0.0839759574168356}
\pgfkeyssetvalue{/bb/j17/switchexpression/Basic/VerificationTime}{1156.8}
\pgfkeyssetvalue{/bb/j17/switchexpression/Basic/SourceSize}{23}
\pgfkeyssetvalue{/bb/j17/switchexpression/Basic/BytecodeSize}{56}
\pgfkeyssetvalue{/bb/j17/switchexpression/Basic/BoogieSize}{135}
\pgfkeyssetvalue{/bb/j17/typeinference/Basic/ConversionTime}{2967.2}
\pgfkeyssetvalue{/bb/j17/typeinference/Basic/ConversionOverhead}{0.0715013882777307}
\pgfkeyssetvalue{/bb/j17/typeinference/Basic/VerificationTime}{1153.8}
\pgfkeyssetvalue{/bb/j17/typeinference/Basic/SourceSize}{29}
\pgfkeyssetvalue{/bb/j17/typeinference/Basic/BytecodeSize}{59}
\pgfkeyssetvalue{/bb/j17/typeinference/Basic/BoogieSize}{116}
\pgfkeyssetvalue{/bb/s/complete/Comparator/ConversionTime}{3435.2}
\pgfkeyssetvalue{/bb/s/complete/Comparator/ConversionOverhead}{0.0930912802678433}
\pgfkeyssetvalue{/bb/s/complete/Comparator/VerificationTime}{1419.2}
\pgfkeyssetvalue{/bb/s/complete/Comparator/SourceSize}{49}
\pgfkeyssetvalue{/bb/s/complete/Comparator/BytecodeSize}{35}
\pgfkeyssetvalue{/bb/s/complete/Comparator/BoogieSize}{237}
\pgfkeyssetvalue{/bb/s/complete/Dice/ConversionTime}{4494.8}
\pgfkeyssetvalue{/bb/s/complete/Dice/ConversionOverhead}{0.1138011962959596}
\pgfkeyssetvalue{/bb/s/complete/Dice/VerificationTime}{1851.8}
\pgfkeyssetvalue{/bb/s/complete/Dice/SourceSize}{38}
\pgfkeyssetvalue{/bb/s/complete/Dice/BytecodeSize}{25}
\pgfkeyssetvalue{/bb/s/complete/Dice/BoogieSize}{223}
\pgfkeyssetvalue{/bb/s/complete/GCD/ConversionTime}{3538.2}
\pgfkeyssetvalue{/bb/s/complete/GCD/ConversionOverhead}{0.082363938085856}
\pgfkeyssetvalue{/bb/s/complete/GCD/VerificationTime}{1308.4}
\pgfkeyssetvalue{/bb/s/complete/GCD/SourceSize}{46}
\pgfkeyssetvalue{/bb/s/complete/GCD/BytecodeSize}{93}
\pgfkeyssetvalue{/bb/s/complete/GCD/BoogieSize}{130}
\pgfkeyssetvalue{/bb/s/complete/LinearSearch/ConversionTime}{3442.0}
\pgfkeyssetvalue{/bb/s/complete/LinearSearch/ConversionOverhead}{0.0982839920537705}
\pgfkeyssetvalue{/bb/s/complete/LinearSearch/VerificationTime}{1264.6}
\pgfkeyssetvalue{/bb/s/complete/LinearSearch/SourceSize}{69}
\pgfkeyssetvalue{/bb/s/complete/LinearSearch/BytecodeSize}{126}
\pgfkeyssetvalue{/bb/s/complete/LinearSearch/BoogieSize}{168}
\pgfkeyssetvalue{/bb/average/ConversionTime}{3094.7441860465115}
\pgfkeyssetvalue{/bb/average/ConversionOverhead}{0.0859427125937027}
\pgfkeyssetvalue{/bb/average/VerificationTime}{1416.0697674418604}
\pgfkeyssetvalue{/bb/average/SourceSize}{44.75}
\pgfkeyssetvalue{/bb/average/BytecodeSize}{80.85}
\pgfkeyssetvalue{/bb/average/BoogieSize}{139.625}
\pgfkeyssetvalue{/bb/total/ConversionTime}{133074.0}
\pgfkeyssetvalue{/bb/total/ConversionOverhead}{3.695536641529216}
\pgfkeyssetvalue{/bb/total/VerificationTime}{60891.0}
\pgfkeyssetvalue{/bb/total/SourceSize}{1790}
\pgfkeyssetvalue{/bb/total/BytecodeSize}{3234}
\pgfkeyssetvalue{/bb/total/BoogieSize}{5585}

%% file: stats.tex
\pgfkeyssetvalue{/bb/stat/cor(bytecode size, Boogie size)}{0.462977352819378}
\pgfkeyssetvalue{/bb/stat/cor(Boogie size, verification time)}{0.363118830922261}
\pgfkeyssetvalue{/bb/stat/Boogie code >= bytecode}{35}
\pgfkeyssetvalue{/bb/stat/Boogie conciseness}{1.81391179634278}

%% file: tables/feature-data.tex
\setlength{\tabcolsep}{1pt}
\begin{tabular}{llclccc}
  \toprule
  \multicolumn{1}{c}{} & & \multicolumn{1}{c}{\textsc{Java}} &  & \multicolumn{3}{c}{\textsc{support}}
  \\
  & \multicolumn{1}{c}{\textsc{feature}} & \multicolumn{1}{c}{\textsc{version}}
                                      & \multicolumn{1}{c}{\textsc{example}} & KeY & Krakatoa & OpenJML
  \\
  \midrule
  G & Generic classes
  & 5
                                     & 
\begin{lstlisting}[language=JavaRecent,style=plain]
class Box<T> {  T value; }
\end{lstlisting} 
                                         &  \suppPart & \suppNO & \suppOK 
  \\
  F & Enhanced \J{for} loop
  & 5
                                      &
\begin{lstlisting}[language=JavaRecent,style=plain]
int[] arr; int res = 0; 
for(int x: arr) res += x;
\end{lstlisting} 
                                         & \suppOK & \suppNO & \suppOK 
  \\
  A & Varargs
  & 5
                                      & 
\begin{lstlisting}[language=JavaRecent,style=plain]
int first(int... values)
{ return values[0]; }
\end{lstlisting} 
                                         & \suppOK & \suppNO & \suppNO 
  \\
  T & Generic type inference
  & 7 
                                      &
\begin{lstlisting}[language=JavaRecent,style=plain]
Box<Integer> b = new Box<>();
\end{lstlisting} 
                                         & \suppNO & \suppNO & \suppNO
  \\
    D & Default methods
    & 8 
                                        &
\begin{lstlisting}[language=JavaRecent,style=plain]
interface PlusMinus extends Plus 
  { default int minus(int x)
      { return plus(-x); } }
\end{lstlisting} 
                                           & \suppOK & \suppNO & \suppOK
    \\
    L & Local type inference
    & 10 
                                        & \ 
\begin{lstlisting}[language=JavaRecent,style=plain]
var b = new Box<Integer>();
\end{lstlisting} 
                                           & \suppNO & \suppNO & \suppNO
    \\
    S & Switch expressions
    & 12 
                                        & \ 
\begin{lstlisting}[language=JavaRecent,style=plain]
System.out.println( switch (day)
  { case 0 -> "Mon";
    default -> "Other"; });
\end{lstlisting} 
                                           & \suppNO & \suppNO & \suppNO
    \\
    Y & Switch expressions with \J{yield}
    & 13
                                        & \ 
\begin{lstlisting}[language=JavaRecent,style=plain]
System.out.println( switch (day)
  { case 0: m++; yield "Mon";
    default: yield "Other"; });
\end{lstlisting} 
                                           & \suppNO & \suppNO & \suppNO
    \\
    P & Pattern matching with \J{instanceof}
    & 14 
                                        & \ 
\begin{lstlisting}[language=JavaRecent,style=plain]
if (obj instanceof String str)
  return str + " is String"; 
\end{lstlisting} 
                                           & \suppNO & \suppNO & \suppNO
    \\
    \bottomrule
\end{tabular}